\pdfoutput=1
\newcommand*{\ATLASLATEXPATH}{}
\documentclass[UKenglish,texlive=2016,backend=bibtex,cernpreprint]{\ATLASLATEXPATH atlasdoc}
 
\usepackage{\ATLASLATEXPATH atlaspackage}
\usepackage[articletitle=true, block=space]{\ATLASLATEXPATH atlasbiblatex}
\usepackage{\ATLASLATEXPATH atlascontribute}
\usepackage[process=true]{\ATLASLATEXPATH atlasphysics}
\graphicspath{{logos/}{figures/}}
 
\usepackage{TOPQ-2016-12-PAPER-defs}
 
\AtlasTitle{Measurement of differential cross-sections of a single top quark produced in association with a $W$ boson at $\sqrt{s}=\SI{13}{\text{TeV}}$ with ATLAS}
 
\author{The ATLAS Collaboration}

\AtlasRefCode{TOPQ-2016-12}
 
\AtlasNote{TOPQ-2016-12}
 
\PreprintIdNumber{CERN-EP-2017-221}

\AtlasJournal{EPJC}
\AtlasJournalRef{\EPJC 78 (2018) 186}
\AtlasDOI{10.1140/epjc/s10052-018-5649-8}
 
\AtlasAbstract{
The differential cross-section for the production of a $W$ boson in association with a top quark is measured for several particle-level observables.
The measurements are performed using ${36.1}\,\text{fb}^{-1}$ of $pp$ collision data collected with the ATLAS detector at the LHC in 2015 and 2016.
Differential cross-sections are measured in a fiducial phase space defined by the presence of two charged leptons and exactly one jet matched to a $b$-hadron,
and are normalised with the fiducial cross-section.
Results are found to be in good agreement with predictions from several Monte Carlo event generators.
}

\AtlasCoverSupportingNote{Measurement of differential cross-sections of $tW$ channel with ATLAS at $\sqs=\SI{13}{\TeV}$}{https://cds.cern.ch/record/2230032}

\AtlasCoverCommentsDeadline{}
 
\AtlasCoverAnalysisTeam{
Ian Brock,
Kevin Finelli,
Carl Suster,
Regina Moles,
Jin Wang,
Rui Zhang}
 
\AtlasCoverEdBoardMember{
Dominic Hirschbuehl,
James Howarth,
Antonio Sidoti (chair)}
 
\AtlasCoverEgroupEditors{atlas-topq-2016-12-editors@cern.ch}
 
\AtlasCoverEgroupEdBoard{atlas-topq-2016-12-editorial-board@cern.ch}
\hypersetup{pdftitle={ATLAS paper},pdfauthor={The ATLAS Collaboration}}
 
\begin{document}
 
\maketitle
 
\tableofcontents
 
\section{Introduction}
\label{sec:intro}
\begin{figure}[t]
\centering
\begin{tabular}{c}
\includegraphics[width=.43\textwidth]{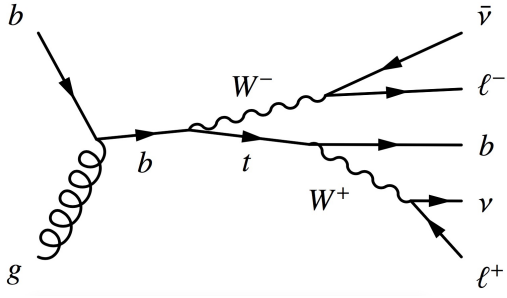}
\end{tabular}
\caption{\label{fig:feyn_wt}
A representative leading-order Feynman diagram for the production of a single top quark in the $tW$ channel and the subsequent leptonic decay of the $W$ boson and semileptonic decay of the top quark.
}
\end{figure}

Single-top-quark production proceeds via three channels through electroweak interactions involving a $Wtb$ vertex at leading order (LO) in the Standard Model (SM): the $t$-channel, the $s$-channel, and production in association with a $W$ boson ($tW$).
The cross-section for each of these channels depends on the relevant Cabibbo-Kobayashi-Maskawa (CKM) matrix element $V_{tb}$ and form factor $f^\mathrm{L}_\mathrm{V}$~\cite{coupling_params1,coupling_params2,coupling_params3} such that the cross-section is proportional to $|f^\mathrm{L}_\mathrm{V}V_{tb}|^{2}$~\cite{coupling_prop1,coupling_prop2}, i.e. depends on the coupling between the $W$ boson, top and $b$ quarks.
The $tW$ channel, represented in \Fig{\ref{fig:feyn_wt}}, has a $pp$ production cross-section at $\sqs=\SI{13}{\TeV}$ of
\ensuremath{\sigma_{\text{theory}}=\xspace{\SI[parse-numbers=false]{71.7\pm1.8\,(\mathrm{scale})\pm{3.4}\,(\mathrm{PDF})}{\fullXsUnit}}}~\cite{Kidonakis:2015nna},
and contributes approximately $24\%$ of the total single-top-quark production rate at \SI{13}{\TeV}.
At the LHC, evidence for this process with \SI{7}{\TeV} collision data was presented by the ATLAS Collaboration~\cite{TOPQ-2011-17} (with a significance of $3.6\sigma$), and by the CMS Collaboration~\cite{CMS-TOP-11-022} (with a significance of $4.0\sigma$).
With \SI{8}{\TeV} collision data, CMS observed the $tW$ channel with a significance of $6.1\sigma$~\cite{CMS-TOP-12-040} while ATLAS observed it with a significance of $7.7\sigma$~\cite{TOPQ-2012-20}.
This analysis extends an ATLAS analysis~\cite{TOPQ-2015-16} which measured the production cross-section with \SI{13}{\TeV} data collected in 2015.
 
Accurate estimates of rates and kinematic distributions of the $tW$ process are difficult at higher orders in $\alphas$ since the process is not well-defined due to quantum interference with the \ttbar{} production process.
A fully consistent theoretical picture can be reached by considering $tW$ and \ttbar{} to be components of the complete $WbWb$ final state in the four flavour scheme~\cite{madgraphtWh}.
In the \ttbar{} process the two $Wb$ systems are produced on the top quark mass shell, and so a proper treatment of this doubly resonant component is important in the study of $tW$ beyond leading order.
Two commonly used approaches are diagram removal (DR) and diagram subtraction (DS)~\cite{dsdr}.
In the DR approach, all next-to-leading order (NLO) diagrams that overlap with the doubly resonant \ttbar{} contributions are removed from the calculation of the $tW$ amplitude, violating gauge invariance.
In the DS approach, a subtraction term is built into the amplitude to cancel out the \ttbar{} component close to the top quark resonance while respecting gauge invariance.
 
This \doctype{} describes differential cross-section measurements in the $tW$ dilepton final state, where events contain two oppositely charged leptons (henceforth ``lepton'' refers to an electron or muon) and two neutrinos.
This channel is chosen because it has a better ratio of signal and \ttbar{} production over other background processes than the single lepton+jets channel, where large $W+$jets backgrounds are relatively difficult to separate from top quark events.
Distributions are unfolded to observables based on stable particles produced in Monte Carlo (MC) simulation.
Measurements are performed in a fiducial phase space, defined by the presence of two charged leptons as well as the presence of exactly one central jet containing $b$-hadrons ($b$-jet) and no other jets.
This requirement on the jet multiplicity is expected to suppress the contribution from \ttbar{} production, where a pair of $b$-jets is more commonly produced, as well as reducing the importance of \ttbar-$tW$ interference effects~\cite{madgraphtWh}.
After applying the reconstruction-level selection of fiducial events (described in \Sect{\ref{sec:events}}) backgrounds from \ttbar{} and other sources are subtracted according to their predicted distributions from MC simulation.
The definition of the fiducial event selection is chosen to match the lepton and jet requirements at reconstruction level.
Exactly two leptons with $\pt>\SI{20}{\GeV}$ and $|\eta|<2.5$ are required, and at least one of the leptons must satisfy $\pt>\SI{27}{\GeV}$.
Exactly one $b$-tagged jet satisfying $\pt>\SI{25}{\GeV}$ and $|\eta|<2.5$ must be present.
No requirement is placed on \met{} or $m_{\ell\ell}$.
A boosted decision tree (BDT) is used to separate the $tW$ signal from the large \ttbar{} background by placing a fixed requirement on the BDT response.
 
Although the top quark and the two $W$ bosons cannot be directly reconstructed due to insufficient kinematic constraints,
one can select a list of observables that are correlated with kinematic properties of $tW$ production and are sensitive to differences in theoretical modelling.
Particle energies and masses are also preferred to projections onto the transverse plane in order to be sensitive to polar angular information while keeping the list of observables as short as possible.
Unfolded distributions are measured for:
 
\begin{itemize}
\item the energy of the $b$-jet, \EJet;
\item the mass of the leading lepton and $b$-jet, \massLepOneJet;
\item the mass of the sub-leading lepton and the $b$-jet, \massLepTwoJet;
\item the energy of the system of the two leptons and $b$-jet, \ELepLepJet;
\item the transverse mass of the leptons, $b$-jet and neutrinos, \massTransLepLepMETJet; and
\item the mass of the two leptons and the $b$-jet, \massLepLepJet.
\end{itemize}
 
The top quark production is probed most directly by \EJet, the only final-state object that can unambiguously be matched to the decay products of the top quark.
The top-quark decay is probed by \massLepOneJet{} and \massLepTwoJet, which are sensitive to angular correlations of decay products due to production spin correlations.
The combined $tW$-system is probed by \ELepLepJet, \massTransLepLepMETJet, and \massLepLepJet.
At reconstruction level, the transverse momenta of the neutrinos in \massTransLepLepMETJet{} are represented by the measured \met{} (reconstructed as described in \Sect{\ref{sec:objects}}).
At particle level the vector summed transverse momenta of simulated neutrinos (selected as defined in \Sect{\ref{sec:objects}}) are used in \massTransLepLepMETJet.
All other quantities for leptons and jets are taken simply from the relevant reconstructed or particle-level objects.
These observables are selected to minimise the bias introduced by the BDT requirement,
as certain observables are highly correlated with the BDT discriminant.
These cannot be effectively unfolded due to shaping effects that the BDT requirement imposes on the overall acceptance, and thus are not considered in this measurement.
The background-subtracted data are unfolded using an iterative procedure~\cite{DAgostini:1994zf} to correct for resolution and acceptance effects, biases, and particles outside the fiducial phase space of the measurement.
The differential cross-sections are normalised with the fiducial cross-section, which cancels out many of the largest uncertainties.
 
\section{ATLAS detector}
The ATLAS detector~\cite{PERF-2007-01} at the LHC covers nearly the entire solid angle
\footnote{
ATLAS uses a right-handed coordinate system with its origin at the nominal interaction point (IP) in the centre of the detector and the $z$-axis along the beam pipe. The $x$-axis points from the IP to the centre of the LHC ring, and the $y$-axis points upward. Cylindrical coordinates $(r,\phi)$ are used in the transverse plane, $\phi$ being the azimuthal angle around the $z$-axis. The pseudorapidity is defined in terms of the polar angle $\theta$ as $\eta=-\ln\tan(\theta/2)$, while the rapidity is defined in terms of particle energies and the $z$-component of particle momenta as $y=(1/2)\ln\left[(E+p_z)/(E-p_z)\right]$.
}
around the collision point, and consists of an inner tracking detector (ID) surrounded by a thin superconducting solenoid producing a \SI{2}{T} axial magnetic field, electromagnetic (EM) and hadronic calorimeters, and an external muon spectrometer (MS).
The ID consists of a high-granularity silicon pixel detector and a silicon microstrip tracker, together providing precision tracking in the pseudorapidity
range $|\eta|<2.5$, complemented by a transition radiation tracker providing tracking and electron identification information for $|\eta|<2.0$.
The innermost pixel layer, the insertable B-layer~\cite{ATLAS-TDR-19}, was added between \RunOne{} and \RunTwo{} of the LHC, at a radius of \SI{33}{mm} around a new, thinner, beam pipe.
A lead liquid-argon (LAr) electromagnetic calorimeter covers the region $|\eta|<3.2$, and hadronic calorimetry is provided by steel\slash{}scintillator tile calorimeters within $|\eta|<1.7$ and copper\slash{}LAr hadronic endcap calorimeters in the range $1.5 < |\eta|< 3.2$.
A LAr forward calorimeter with copper and tungsten absorbers covers the range $3.1 < |\eta| <4.9$.
The MS consists of precision tracking chambers covering the region $|\eta|<2.7$, and separate trigger chambers covering $|\eta|<2.4$.
A two-level trigger system~\cite{TRIG-2016-01}, using a custom hardware level followed by a software-based level, selects from the \SI{40}{MHz} of collisions a maximum of around \SI{1}{kHz} of events for offline storage.
 
\section{Data and Monte Carlo samples}
The data events analysed in this \doctype{} correspond to an integrated luminosity of \lumi{} collected from the operation of the LHC in 2015 and 2016 at $\sqs=\SI{13}{\TeV}$ with a bunch spacing of \SI{25}{ns} and an average number of collisions per bunch crossing $\langle\mu\rangle$ of around 23.
They are required to be recorded in periods where all detector systems are flagged as operating normally.
 
Monte Carlo simulated samples are used to estimate the efficiency to select signal and background events, train and test the BDT, estimate the migration of observables from particle level to reconstruction level, estimate systematic uncertainties, and validate the analysis tools.
The nominal samples, used for estimating the central values for efficiencies and background templates, were simulated with a full ATLAS detector simulation~\cite{SOFT-2010-01} implemented in \GEANT~4~\cite{geant4}.
Many of the samples used in the estimation of systematic uncertainties were instead produced using \textsc{Atlfast2}~\cite{ATL-PHYS-PUB-2010-013}, in which a parameterised detector simulation is used for the calorimeter responses.
Pile-up (additional $pp$ collisions in the same or a nearby bunch crossing) is included in the simulation by overlaying collisions with the soft QCD processes from \PYTHIA~8.186~\cite{pythia81} using a set of tuned parameters called the A2 tune~\cite{ATL-PHYS-PUB-2012-003} and the MSTW2008LO parton distribution function (PDF) set~\cite{mstw}.
Events were generated with a predefined distribution of the expected number of interactions per bunch crossing, then reweighted to match the actual observed data conditions.
In all MC samples and fixed-order calculations used for this analysis the top quark mass $m_{\text{t}}$ is set to \SI{172.5}{\GeV} and the $W\to \ell \nu$ branching ratio is set to $0.108$ per lepton flavour.
The \textsc{EvtGen}~v1.2.0 program~\cite{evtgen} was used to simulate properties of the bottom and charmed hadron decays except for samples generated with \SHERPA, which uses internal modules.
 
The nominal $tW$ event samples~\cite{ATL-PHYS-PUB-2016-004} were produced using the \POWHEGBOX~v1~\cite{Nason:2004rx,Frixione:2007vw,Alioli:2010xd,Re:2010bp,Campbell:2014kua} event generator with the CT10 PDF set~\cite{ct10} in the matrix-element calculations.
The parton shower, hadronisation, and underlying event were simulated using \PYTHIAV{6.428}~\cite{pythia6} with the CTEQ6L1 PDF set~\cite{cteq6} and the corresponding Perugia~2012 (P2012) tune~\cite{perugia}.
The DR scheme~\cite{dsdr} was employed to handle the interference between $tW$ and \ttbar, and was applied to the $tW$ sample.
For comparing MC predictions to data, the predicted $tW$ cross-section at $\sqs=\SI{13}{\TeV}$ is scaled by a $K$-factor and set to the NLO value with next-to-next-to-leading logarithmic (NNLL) soft-gluon corrections:
\ensuremath{\sigma_{\text{theory}}=\xspace{\SI[parse-numbers=false]{71.7\pm1.8\,(\mathrm{scale})\pm{3.4}\,(\mathrm{PDF})}{\fullXsUnit}}}~\cite{Kidonakis:2015nna}.
The first uncertainty accounts for the renormalisation and factorisation scale variations (from $0.5$ to $2$ times $m_{\text{t}}$), while the second uncertainty originates from uncertainties in the MSTW2008 NLO PDF sets.
 
Additional $tW$ samples were generated to estimate systematic uncertainties in the modelling of the signal process.
An alternative $tW$ sample was generated using the DS scheme instead of DR\@.
A $tW$ sample generated with \MGMCatNLO~v2.2.2~\cite{Alwall:2014hca} (instead of the \POWHEGBOX) interfaced with \HERWIGpp~2.7.1~\cite{herwig} and processed through the \textsc{Atlfast2} fast simulation is used to estimate uncertainties associated with the modelling of the NLO matrix-element event generator.
A sample generated with \POWHEGBOX{} interfaced with \HERWIGpp{} (instead of \PYTHIAV{6}) is used to estimate uncertainties associated with the parton shower, hadronisation, and underlying-event models.
This sample is also compared with the previously mentioned \MGMCatNLO{} sample to estimate a matrix-element event generator uncertainty with a consistent parton shower event generator.
In both cases, the UE-EE-5 tune of~\Ref{\cite{ATL-PHYS-PUB-2014-021}} was used for the underlying event.
Finally, in order to estimate uncertainties arising from additional QCD radiation in the $tW$ events, a pair of samples were generated with \POWHEGBOX{} interfaced with \PYTHIAV{6} using \textsc{Atlfast2} and the P2012 tune with higher and lower radiation relative to the nominal set, together with varied renormalisation and factorisation scales.
In order to avoid comparing two different detector response models when estimating systematic uncertainties, another version of the nominal \POWHEGBOX{} with \PYTHIAV{6} sample was also produced with \textsc{Atlfast2}.
 
The nominal \ttbar{} event sample~\cite{ATL-PHYS-PUB-2016-004} was produced using the \POWHEGBOX~v2~\cite{Nason:2004rx,Frixione:2007vw,Alioli:2010xd,Re:2010bp,Campbell:2014kua} event generator with the CT10 PDF set~\cite{ct10} in the matrix-element calculations.
The parton shower, hadronisation, and underlying event were simulated using \PYTHIAV{6.428}~\cite{pythia6} with the CTEQ6L1 PDF set~\cite{cteq6} and the corresponding Perugia~2012 (P2012) tune~\cite{perugia}.
The renormalisation and factorisation scales are set to $m_{\text{t}}$ for the $tW$ process and to $\sqrt{m_{\text{t}}^2 + {\pt(t)}^2}$ for the \ttbar{} process, and the $h_\text{damp}$ resummation damping factor is set to equal the mass of the top quark.
 
Additional \ttbar{} samples were generated to estimate systematic uncertainties.
Like the additional $tW$ samples, these are used to estimate the uncertainties associated with the matrix-element event generator (a sample produced using \textsc{Atlfast2} fast simulation with \MGMCatNLO~v2.2.2 interfaced with \HERWIGpp~2.7.1), parton shower and hadronisation models (a sample produced using \textsc{Atlfast2} with \POWHEGBOX{} interfaced with \HERWIGpp~2.7.1) and additional QCD radiation.
To estimate uncertainties on additional QCD radiation in \ttbar, a pair of samples is produced using full simulation with the varied sets of P2012 parameters for higher and lower radiation, as well as with varied renormalisation and factorisation scales.
In these samples the resummation damping factor $h_\text{damp}$ is doubled in the case of higher radiation.
The \ttbar{} cross-section is set to $\sigma_{\ttbar} = \SI[parse-numbers=false]{831.8\,^{+19.8}_{-29.2}\,(\mathrm{scale})\,\pm35.1\,(\mathrm{PDF} + \alphas)}{pb}$ as calculated with the \textsc{Top++}~2.0 program to NNLO, including soft-gluon resummation to NNLL~\cite{top++}.
The first uncertainty comes from the independent variation of the factorisation and renormalisation scales, $\mu_{\mathrm{F}}$ and $\mu_{\mathrm{R}}$, while the second one is associated with variations in the PDF and $\alphas$, following the PDF4LHC prescription with the MSTW2008 $68\%$ CL NNLO, CT10 NNLO and NNPDF2.3 5f FFN PDF sets~\cite{Botje:2011sn,Martin:2009bu,Gao:2013xoa,nnpdf2.3}.
 
Samples used to model the \Zjets{} background~\cite{ATL-PHYS-PUB-2016-003} were simulated with \SHERPAV{2.2.1}~\cite{sherpa}.
In these, the matrix element is calculated for up to two partons at NLO and four partons at LO using \textsc{Comix}~\cite{comix} and \textsc{OpenLoops}~\cite{openloops}, and merged with the \SHERPA{} parton shower~\cite{csshower} using the \textsc{ME+PS@NLO} prescription~\cite{mepsnlo}.
The NNPDF3.0 NNLO PDF set~\cite{nnpdf3.0} was used in conjunction with \SHERPA{} parton shower tuning, with a generator-level cut-off on the dilepton invariant mass of $\mll>\SI{40}{\GeV}$ applied.
The \Zjets{} events are normalised using NNLO cross-sections computed with FEWZ~\cite{Gavin:2010az}.
 
Diboson processes with four charged leptons, three charged leptons and one neutrino, or two charged leptons and two neutrinos~\cite{ATL-PHYS-PUB-2016-002} were simulated using the \SHERPAV{2.1.1} event generator.
The matrix elements contain all diagrams with four electroweak vertices.
NLO calculations were used for the purely leptonic final states as well as for final states with two or four charged leptons plus one additional parton.
For other final states with up to three additional partons, the LO calculations of \textsc{Comix} and \textsc{OpenLoops} were used.
Their outputs were combined with the \SHERPA{} parton shower using the ME+PS@NLO prescription~\cite{mepsnlo}.
The CT10 PDF set with dedicated parton shower tuning was used.
The cross-sections provided by the event generator (which are already at NLO) were used for diboson processes.
 
\section{Object reconstruction}
\label{sec:objects}
Electron candidates are reconstructed from energy deposits in the EM calorimeter associated with ID tracks~\cite{TRIG-2016-01}.
The deposits are required to be in the $|\eta|<2.47$ region, with the transition region between the barrel and endcap EM calorimeters, $1.37<|\eta|<1.52$, excluded.
The candidate electrons are required to have a transverse momentum of $\pt>\SI{20}{\GeV}$.
Further requirements on the electromagnetic shower shape, ratio of calorimeter energy to tracker momentum, and other variables are combined into a likelihood-based discriminant~\cite{ATLAS-CONF-2016-024}, with signal electron efficiencies measured to be at least 85\%, increasing for higher \pt.
Candidate electrons also must satisfy requirements on the distance from the ID track to the beamline or to the reconstructed primary vertex in the event, which is identified as the vertex with the largest summed $\pt^2$ of associated tracks.
The transverse impact parameter with respect to the beamline, $d_0$, must satisfy $|d_0|/\sigma_{d_{0}} < 5$, where $\sigma_{d_0}$ is the uncertainty in $d_0$.
The longitudinal impact parameter, $z_0$, must satisfy $|\Delta z_0 \sin \theta |<\SI{0.5}{mm}$, where $\Delta z_0$ is the longitudinal distance from the primary vertex along the beamline and $\theta$ is the angle of the track to the beamline.
Furthermore, electrons must satisfy isolation requirements based on ID tracks and topological clusters in the calorimeter~\cite{PERF-2014-07}, designed to achieve an isolation efficiency of $90\%$~($99\%$) for $\pt=\SI[parse-numbers=false]{25(60)}{\GeV}$.
 
Muon candidates are identified by matching MS tracks with ID tracks~\cite{PERF-2015-10}.
The candidates must satisfy requirements on hits in the MS and on the compatibility of ID and MS momentum measurements to remove fake muon signatures.
Furthermore, they must have $\pt>\SI{20}{\GeV}$ as well as $|\eta|<2.5$ to ensure they are within coverage of the ID\@.
Candidate muons must satisfy the following requirements on the distance from the combined ID and MS track to the beamline or primary vertex: the transverse impact parameter significance must satisfy $|d_0|/\sigma_{d_{0}} < 3$, and the longitudinal impact parameter must satisfy $|\Delta z_0 \sin \theta |<\SI{0.5}{mm}$, where $d_0$ and $z_0$ are defined as above for electrons.
An isolation requirement based on ID tracks and topological clusters in the calorimeter is imposed, which targets an isolation efficiency of $90\%$~($99\%$) for $\pt=\SI[parse-numbers=false]{25(60)}{\GeV}$.
 
Jets are reconstructed from topological clusters of energy deposited in the calorimeter~\cite{PERF-2014-07} using the anti-$k_t$ algorithm~\cite{antikt1} with a radius parameter of \SI{0.4} implemented in the FastJet package~\cite{fastjet}.
Their energies are corrected to account for pile-up and calibrated using a \pt- and $\eta$-dependent correction derived from \RunTwo{} data~\cite{ATL-PHYS-PUB-2015-015}.
They are required to have $\pt>\SI{25}{\GeV}$ and $|\eta|<2.5$.
To suppress pile-up, a discriminant called the jet-vertex-tagger is constructed using a two-dimensional likelihood method~\cite{ATLAS-CONF-2014-018}.
For jets with $\pt<\SI{60}{\GeV}$ and $|\eta| < 2.4$, a jet-vertex-tagger requirement corresponding to a $92\%$ efficiency while rejecting $98\%$ of jets from pile-up and noise is imposed.

The tagging of $b$-jets uses a multivariate discriminant which exploits the long lifetime of $b$-hadrons and large invariant mass of their decay products relative to $c$-hadrons and unstable light hadrons~\cite{PERF-2012-04,ATL-PHYS-PUB-2016-012}.
The discriminant is calibrated to achieve a $77\%$ \btag{} efficiency and a rejection factor of about 4.5 against jets containing charm quarks ($c$-jets) and 140 against light-quark and gluon jets in a sample of simulated \ttbar{} events.
The jet tagging efficiency in simulation is corrected to the efficiency in data~\cite{ATL-PHYS-PUB-2015-039}.

The missing transverse momentum vector is calculated as the negative vectorial sum of the transverse momenta of particles in the event.
Its magnitude, \met, is a measure of the transverse momentum imbalance, primarily due to neutrinos that escape detection.
In addition to the identified jets, electrons and muons, a track-based soft term is included in the \met{} calculation by considering tracks associated with the hard-scattering vertex in the event which are not also associated with an identified jet, electron, or muon~\cite{ATL-PHYS-PUB-2015-023,ATL-PHYS-PUB-2015-027}.

To avoid cases where the detector response to a single physical object is reconstructed as two separate final-state objects, several steps are followed to remove such overlaps.
First, identified muons that deposit energy in the calorimeter and share a track with an electron are removed, followed by the removal of any remaining electrons sharing a track with a muon.
This step is designed to avoid cases where a muon mimics an electron through radiation of a hard photon.
Next, the jet closest to each electron within a $y$--$\phi$ cone of size $\DeltaR\equiv\sqrt{{(\Delta y)}^2 + {(\Delta\phi)}^2} = 0.2$ is removed to reduce the proportion of electrons being reconstructed as jets.
Next, electrons with a distance $\DeltaR < 0.4$ from any of the remaining jets are removed to reduce backgrounds from non-prompt, non-isolated electrons originating from heavy-flavour hadron decays.
Jets with fewer than three tracks and distance $\DeltaR < 0.2$ from a muon are then removed to reduce the number of jet fakes from muons depositing energy in the calorimeters.
Finally, muons with a distance $\DeltaR < 0.4$ from any of the surviving jets are removed to avoid contamination due to non-prompt muons from heavy-flavour hadron decays.

Definitions of particle-level objects in MC simulation are based on stable ($c \tau>\SI{10}{mm}$) outgoing particles~\cite{ATL-PHYS-PUB-2015-013}.
Particle-level prompt charged leptons and neutrinos that arise from decays of $W$ bosons or $Z$ bosons are accepted.
The charged leptons are then dressed with nearby photons, considering all photons that satisfy $\DeltaR(\ell,\gamma) < 0.1$ and do not originate from hadrons, adding the four-momenta of all selected photons to the bare lepton to obtain the dressed lepton four-momentum.
Particle-level jets are built from all remaining stable particles in the event after excluding leptons and the photons used to dress the leptons, clustering them using the anti-$k_t$ algorithm with $R=0.4$.
Particle-level jet $b$-tagging is performed by checking the jets for any associated $b$-hadron with $\pt>\SI{5}{\GeV}$.
This association is achieved by reclustering jets with $b$-hadrons included in the input list of particles, but with their \pt{} scaled down to negligibly small values.
Jets containing $b$-hadrons after this reclustering are considered to be associated to a $b$-hadron.
 
\section{Event selection}
\label{sec:events}
Events passing the reconstruction-level selection are required to have at least one interaction vertex, to pass a single-electron or single-muon trigger, and to contain at least one jet with $\pt>\SI{25}{\GeV}$.
Single-lepton triggers used in this analysis are designed to select events containing a well-identified charged lepton with high transverse momentum~\cite{TRIG-2016-01}.
They require a \pt{} of at least \SI{20}{\GeV} (\SI{26}{\GeV}) for muons and \SI{24}{\GeV} (\SI{26}{\GeV}) for electrons for the 2015 (2016) data set, and also have requirements on the lepton quality and isolation.
These are complemented by triggers with higher \pt{} thresholds and relaxed isolation and identification requirements to ensure maximum efficiency at higher lepton \pt.
 
Events are required to contain exactly two oppositely charged leptons with $\pt>\SI{20}{\GeV}$; events with a third charged lepton with $\pt>\SI{20}{\GeV}$ are rejected.
At least one lepton must have $\pt>\SI{27}{\GeV}$, and at least one of the selected electrons (muons) must be matched within a $\DeltaR$ cone of size $0.07$ ($0.1$) to the electron (muon) selected online by the corresponding trigger.
 
In simulated events, information recorded by the event generator is used to identify events in which any selected lepton does not originate promptly from the hard-scatter process.
These non-prompt or fake leptons arise from processes such as the decay of a heavy-flavour hadron, photon conversion or hadron misidentification, and are identified when the electron or muon does not originate from the decay of a $W$ or $Z$ boson (or a $\tau$ lepton itself originating from a $W$ or $Z$).
Events with a selected lepton which is non-prompt or fake are themselves labelled as fake and, regardless of whether they are $tW$ fake events or fake events from other sources, they are treated as a contribution to the background.
 
After this selection has been made, a further set of requirements is imposed with the aim of reducing the contribution from the \Zjets, diboson and fake-lepton backgrounds.
The samples consist almost entirely of $tW$ signal and \ttbar{} background, which are subsequently separated by the BDT discriminant.
Events in which the two leptons have the same flavour and an invariant mass consistent with a $Z$ boson ($81<m_{\ell\ell}<\SI{101}{\GeV}$) are vetoed, as well as those with an invariant mass $m_{\ell\ell}<\SI{40}{\GeV}$.
Further requirements placed on \met{} and $m_{\ell\ell}$ depend on the flavour of the selected leptons.
Events with different-flavour leptons contain backgrounds from $Z\to\tau\tau$, and are required to have $\met>\SI{20}{\GeV}$, with the requirement raised to $\met>\SI{50}{\GeV}$ when the dilepton invariant mass satisfies $\mll<\SI{80}{\GeV}$.
All events with same-flavour leptons, which contain backgrounds from $Z\to ee$ and $Z\to\mu\mu$, must satisfy $\met>\SI{40}{\GeV}$.
For same-flavour leptons, the \Zjets{} background is concentrated in a region of the \mbox{\mll--\met{} plane} corresponding to values of \mll{} near the $Z$ mass, and towards low values of \met.
Therefore, a selection in \met{} and \mll{} is used to remove these backgrounds: events with $\SI{40}{\GeV}<\mll<\SI{81}{\GeV}$ are required to satisfy
$\met{} > 1.25\times\mll$ while events with $\mll>\SI{101}{\GeV}$ are required to satisfy $\met>\SI{300}{\GeV} - 2\times\mll$.
 
\begin{figure}[tbp]
\centering
\includegraphics[width=.50\textwidth]{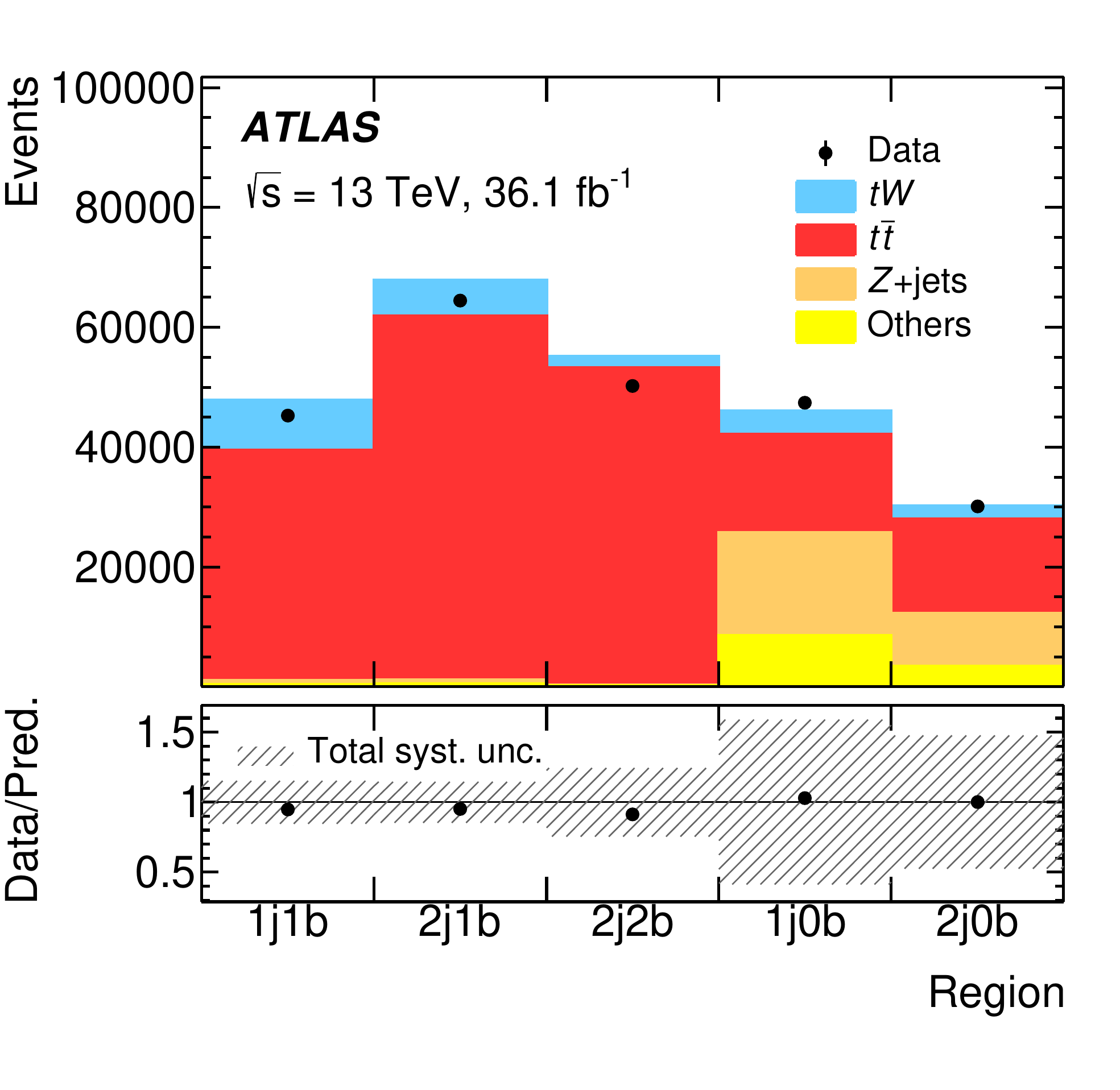}
\caption{\label{fig:regions}
Expected event yields for signal and backgrounds with their total systematic uncertainty (discussed in \Sect{\ref{sec:systematics}}) and the number of observed events in data shown in the signal region (labelled 1j1b) and the four additional regions (labelled 2j1b, 2j2b, 1j0b and 2j0b, based on the number of selected jets and $b$-tagged jets).
``Others'' includes diboson and fake-lepton backgrounds.
The signal and backgrounds are normalised to their theoretical predictions, and the error bands in the lower panel represent the total systematic uncertainties which are used in this analysis.
The upper panel gives the yields in number of events per bin, while the lower panel gives the ratios of the numbers of observed events to the total prediction in each bin.
}
\end{figure}

\begin{figure}
\centering
\begin{tabular}{cc}
\includegraphics[width=.43\textwidth]{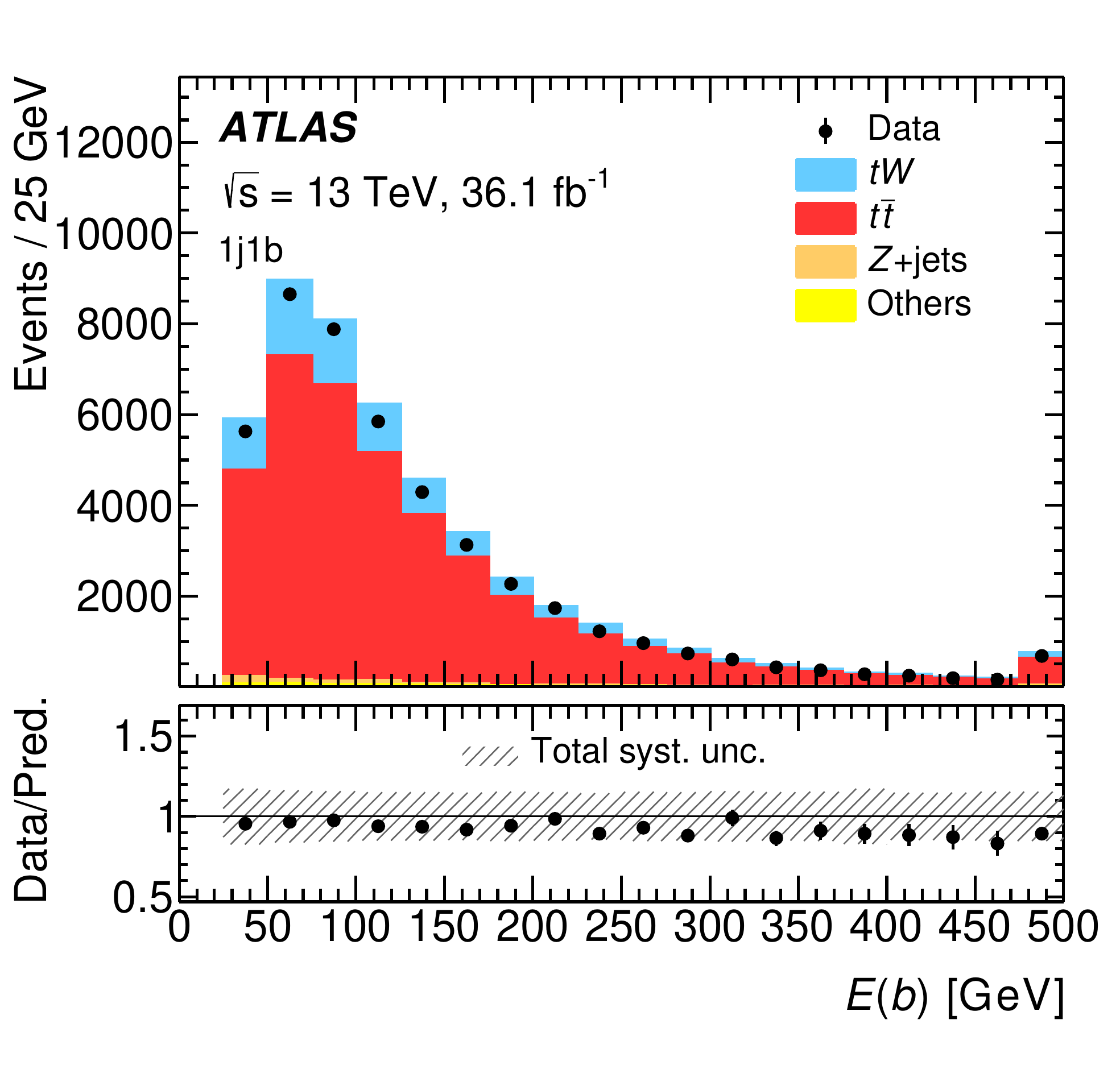} &
\includegraphics[width=.43\textwidth]{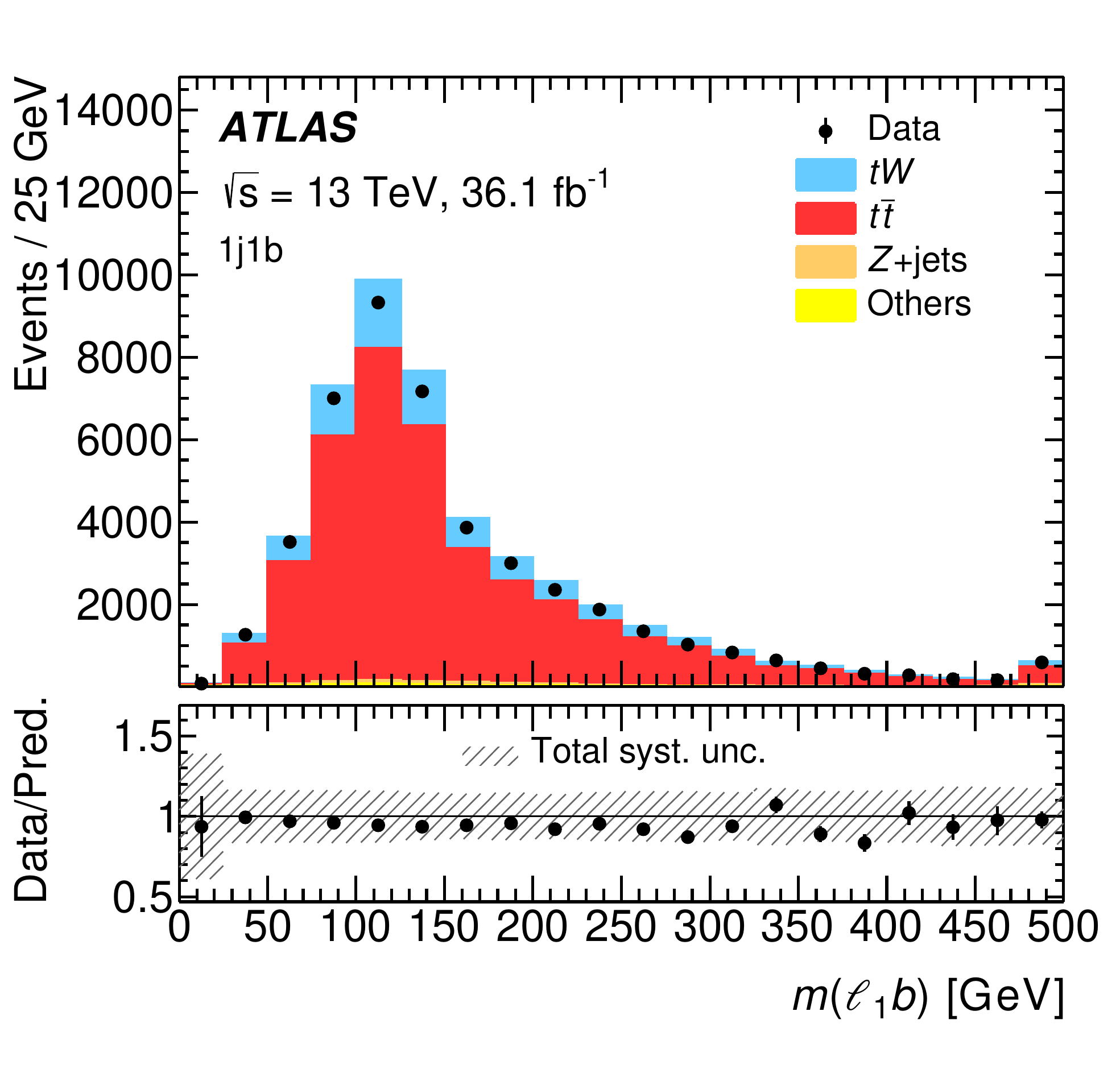} \\
\includegraphics[width=.43\textwidth]{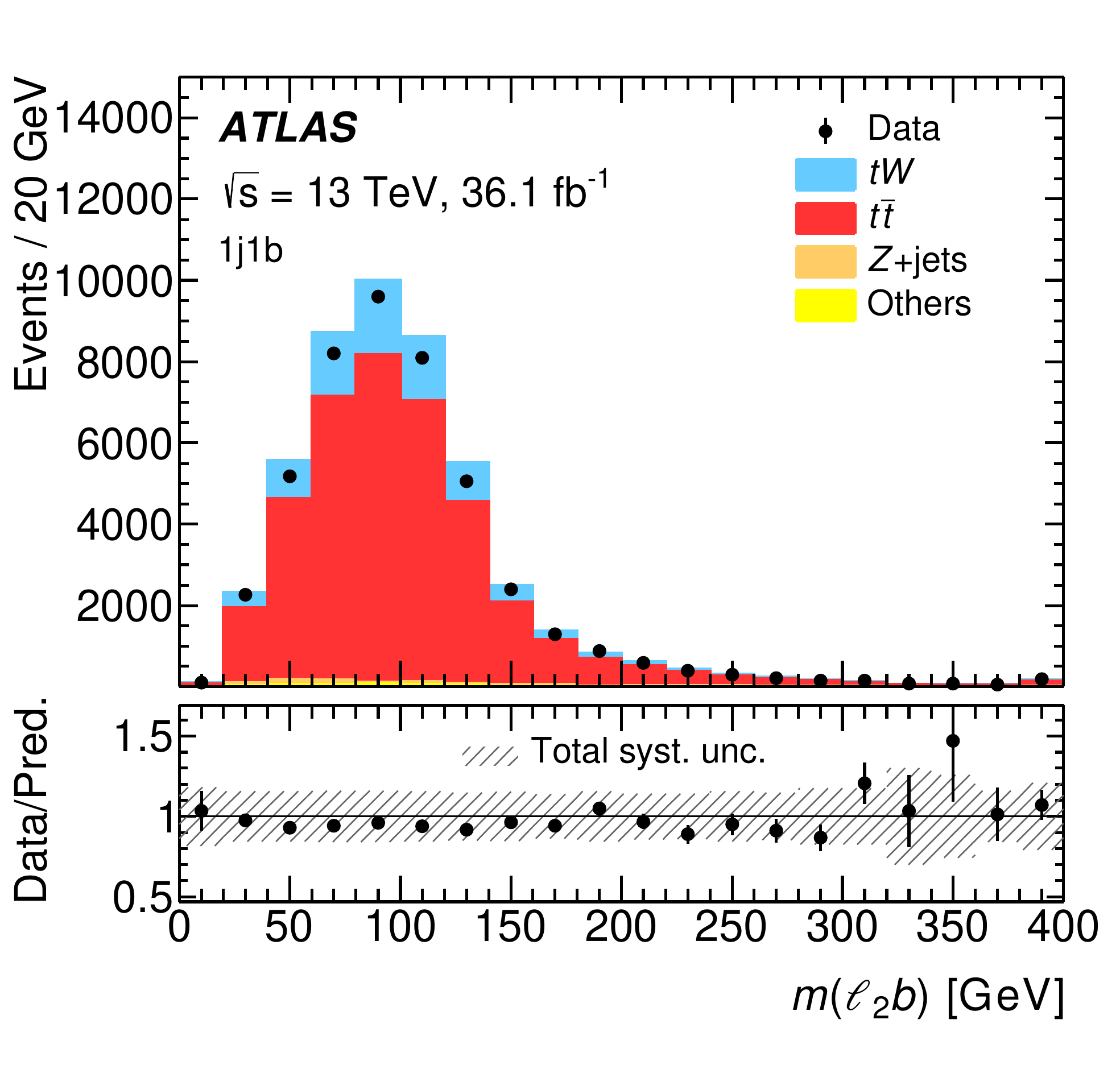} &
\includegraphics[width=.43\textwidth]{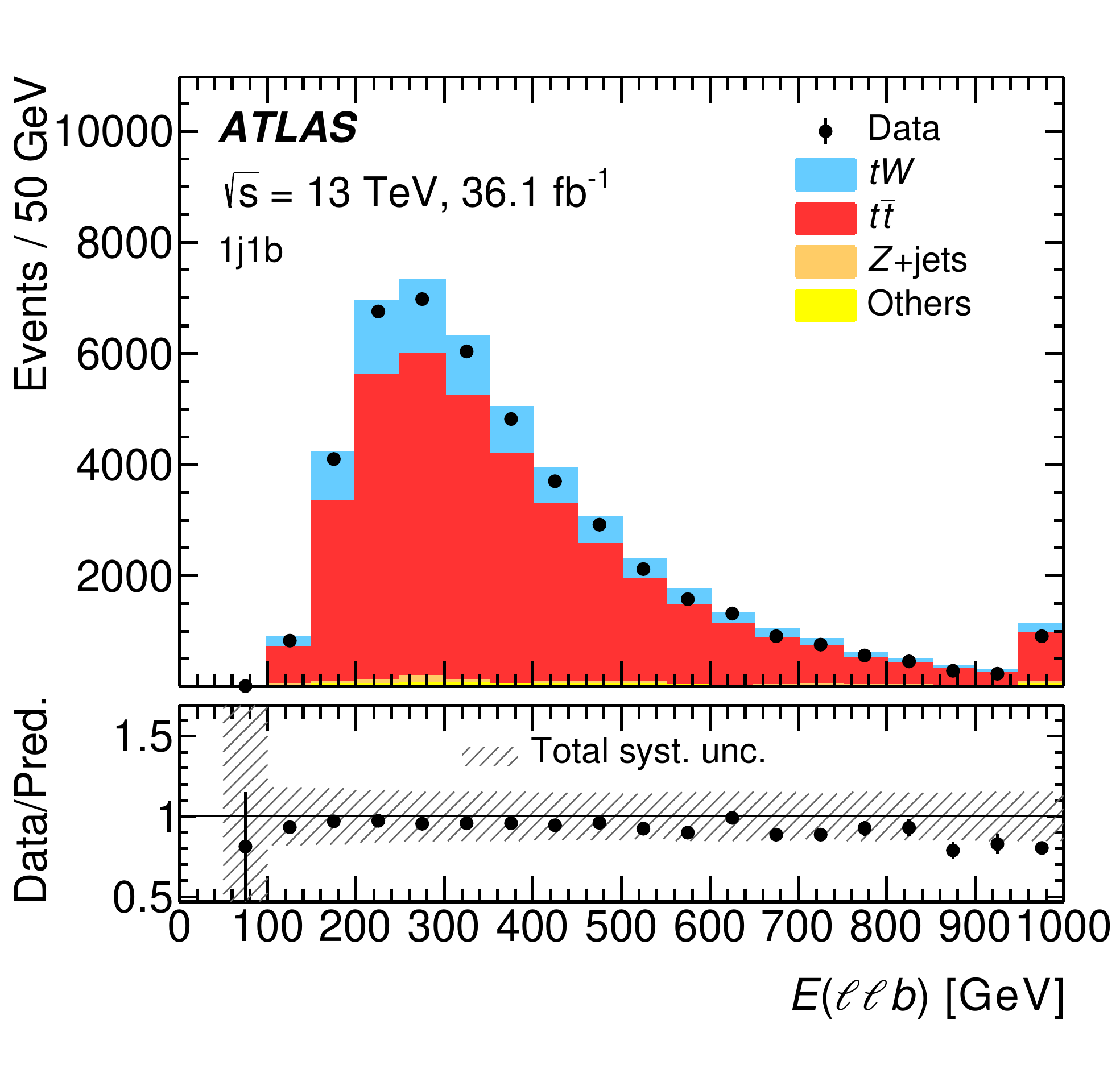} \\
\includegraphics[width=.43\textwidth]{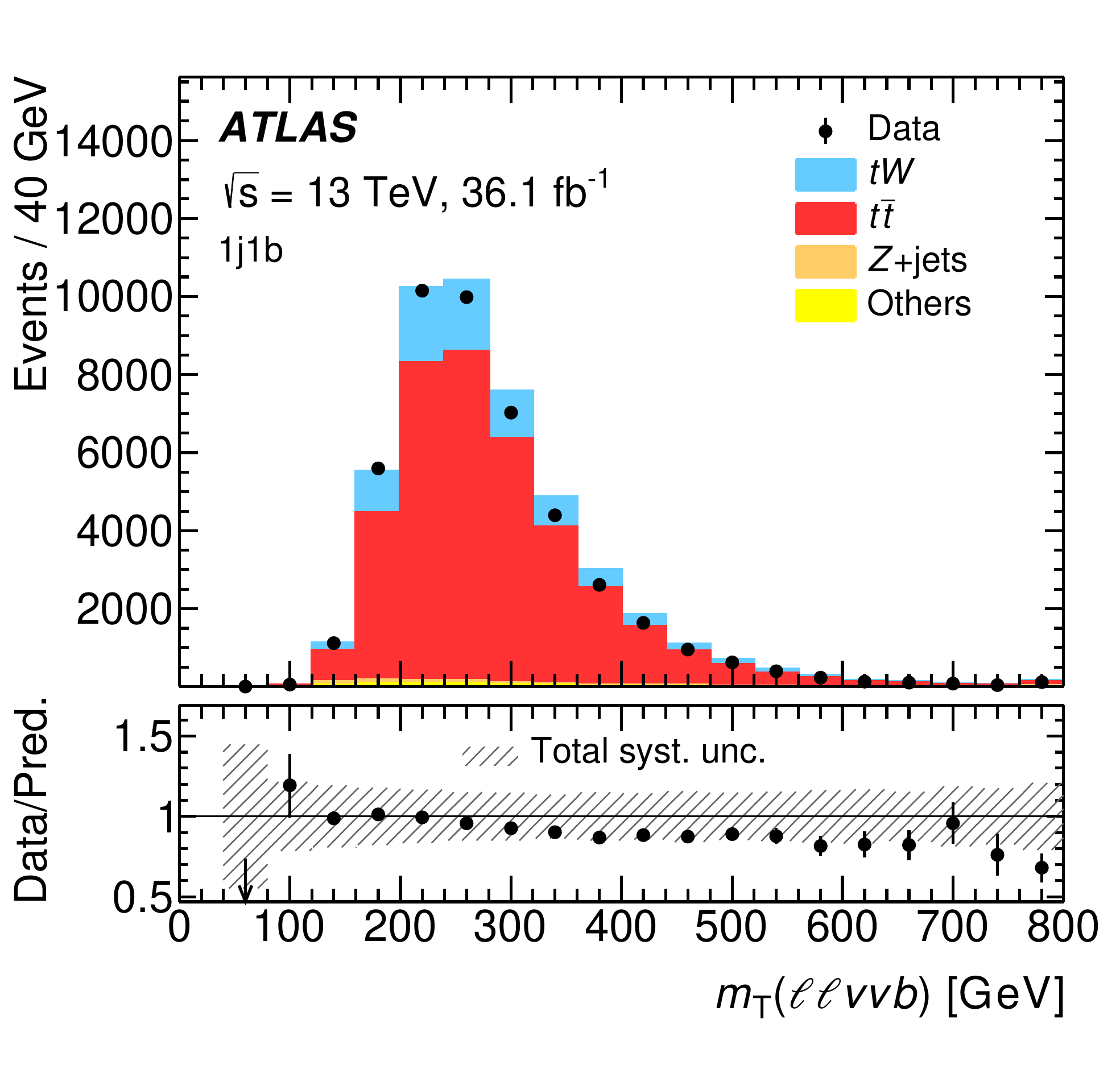} &
\includegraphics[width=.43\textwidth]{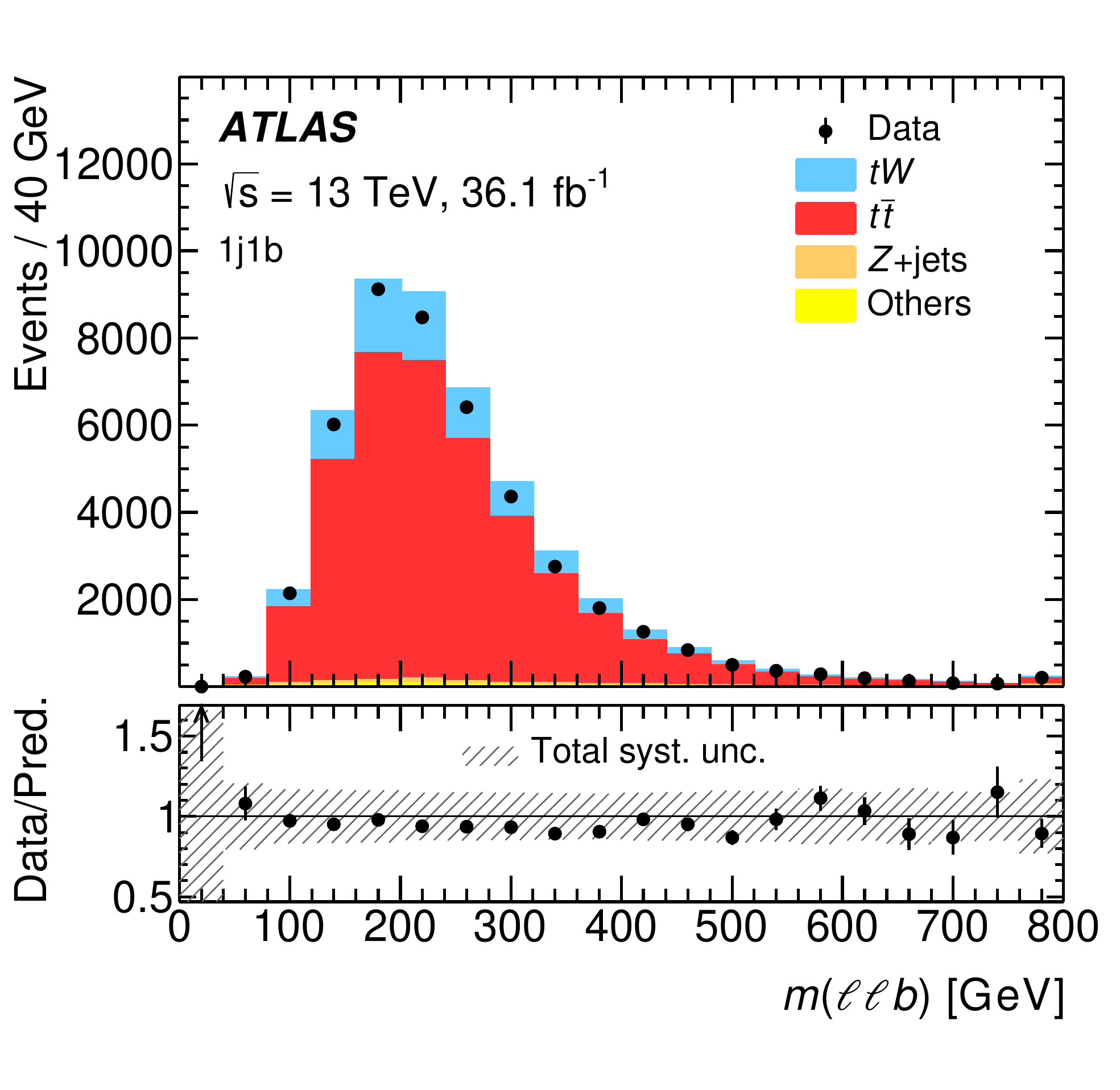} \\
\end{tabular}
\caption{\label{fig:variables}
Distributions of the observables chosen to be unfolded after selection at the reconstruction level but before applying the BDT selection.
The signal and backgrounds are normalised to their theoretical predictions, and the error bands represent the total systematic uncertainties in the MC predictions.
The last bin of each distribution contains overflow events.
The panels give the yields in number of events, and the ratios of the numbers of observed events to the total prediction in each bin.
}
\end{figure}

Finally, events are required to have exactly one jet which is $b$-tagged.
For validation of the signal and background models, additional regions are also defined according to the number of jets and the number of $b$-tagged jets, but are not used in the differential cross-section measurement, primarily due to the lower signal purity in these regions.
These regions are labelled by the number $n$ of selected jets and the number $m$ of selected $b$-tagged jets as $n$j$m$b (for example the 2j1b region consists of events with 2 selected jets of which 1 is $b$-tagged), and show good agreement between data and predictions.
The event yields for signal and backgrounds with their total systematic uncertainties, as well as the number of observed events in the data in the signal and validation regions are shown in \Fig{\ref{fig:regions}}, and the yields in the signal region are shown in \Tab{\ref{tab:yields}}.
Distributions of the events passing these requirements are shown in \Fig{\ref{fig:variables}} at reconstruction level.
Most of the predictions agree well with data within the systematic errors, which are highly correlated bin-to-bin due to the dominance of a small number of sources of large normalisation uncertainties.
The distribution of \massTransLepLepMETJet{}, which shows a slope in the ratio of data to prediction, has a $p$-value of $2$--$4$\% for the predictions to describe the observed distribution after taking bin-to-bin correlations into account.
 
\begin{DIFnomarkup}
\begin{table}
\renewcommand{\arraystretch}{1.1}
\sisetup{group-digits=integer, group-minimum-digits=4}
\centering
\caption{\label{tab:yields}
Predicted and observed yields in the 1j1b signal region before
and after the application of the BDT requirement.}
\begin{tabular}{l
S[table-format=5.0]@{$\,\pm\,$}S[table-format=4.0]
S[table-format=7.0]@{$\,\pm\,$}S[table-format=4.0]@{\hspace*{2em}}
}
\toprule
Process & \multicolumn{2}{c}{Events}              & \multicolumn{2}{c}{Events} \\
& \multicolumn{2}{c}{\mbox{}}             & \multicolumn{2}{c}{BDT response $> 0.3$} \\
\midrule
$tW$    & 8300 & 1400     &  1970 & 560        \\
\ttbar{}  & 38400 & 6600   &  3400 & 1300       \\
\Zjets{}  &  620 & 310      &  159  & 80         \\
Diboson &   230 & 58      &  81   & 20         \\
Fakes   &   220 & 220     &  19   & 19         \\
\midrule
Predicted& 47800 & 7300  &  5600 & 1700     \\
Observed & \multicolumn{1}{S[table-format=5.0]@{\phantom{$\,\pm\,$}}}{45273} &  &  \multicolumn{1}{S[table-format=7.0]@{\phantom{$\,\pm\,$}}}{5043} & \\
\bottomrule
\end{tabular}
\end{table}
\end{DIFnomarkup}
 
\FloatBarrier{}
 
\section{Separation of \texorpdfstring{$tW$}{tW} signal from \texorpdfstring{\ttbar{}}{tt} background}
To separate $tW$ signal events from background \ttbar{} events, a BDT technique~\cite{bdt} is used to combine several observables into a single discriminant.
In this analysis, the BDT implementation is provided by the \textsc{TMVA} package~\cite{tmva}, using the GradientBoost algorithm.
The approach is based on the BDT developed for the inclusive cross-section measurement in \Ref{\cite{TOPQ-2015-16}}.
 
The BDT is optimised by using the sum of the nominal $tW$ MC sample, the alternative $tW$ MC sample with the diagram subtraction scheme and the nominal \ttbar{} MC sample; for each sample, half of the events are used for training while the other half is reserved for testing.
A large list of variables is prepared to serve as inputs to the BDT\@.
An optimisation procedure is then carried out to select a subset of input variables and a set of BDT parameters (such as the number of trees in the ensemble and the maximum depth of the individual decision trees).
The optimisation is designed to provide the best separation between the $tW$ signal and the \ttbar{} background while avoiding sensitivity to statistical fluctuations in the training sample.
 
The variables considered are derived from the kinematic properties of subsets of the selected physics objects defined in \Sect{\ref{sec:objects}} for each event.
For a set of objects $o_1 \ldots o_n$:
$\pt(o_1 \ldots o_n)$ is the transverse momentum of vector sums of various subsets;
$\sum{\ET}$ is the scalar sum of the transverse momenta of all objects which contribute to the \met{} calculation;
$\eta(o_1 \ldots o_n)$ is the pseudorapidity of vector sums of various subsets;
$m(o_1 \ldots o_n)$ is the invariant mass of various subsets.
For vector sums of two systems of objects $s_{1}$ and $s_{2}$:
$\Delta\pt(s_1, s_2)$ is the \pt{} difference; and
$C(s_1 s_2)$ is the ratio of the scalar sum of \pt{} to the sum of energy, called the centrality.
 
\begin{figure}
\centering
\includegraphics[width=.50\textwidth]{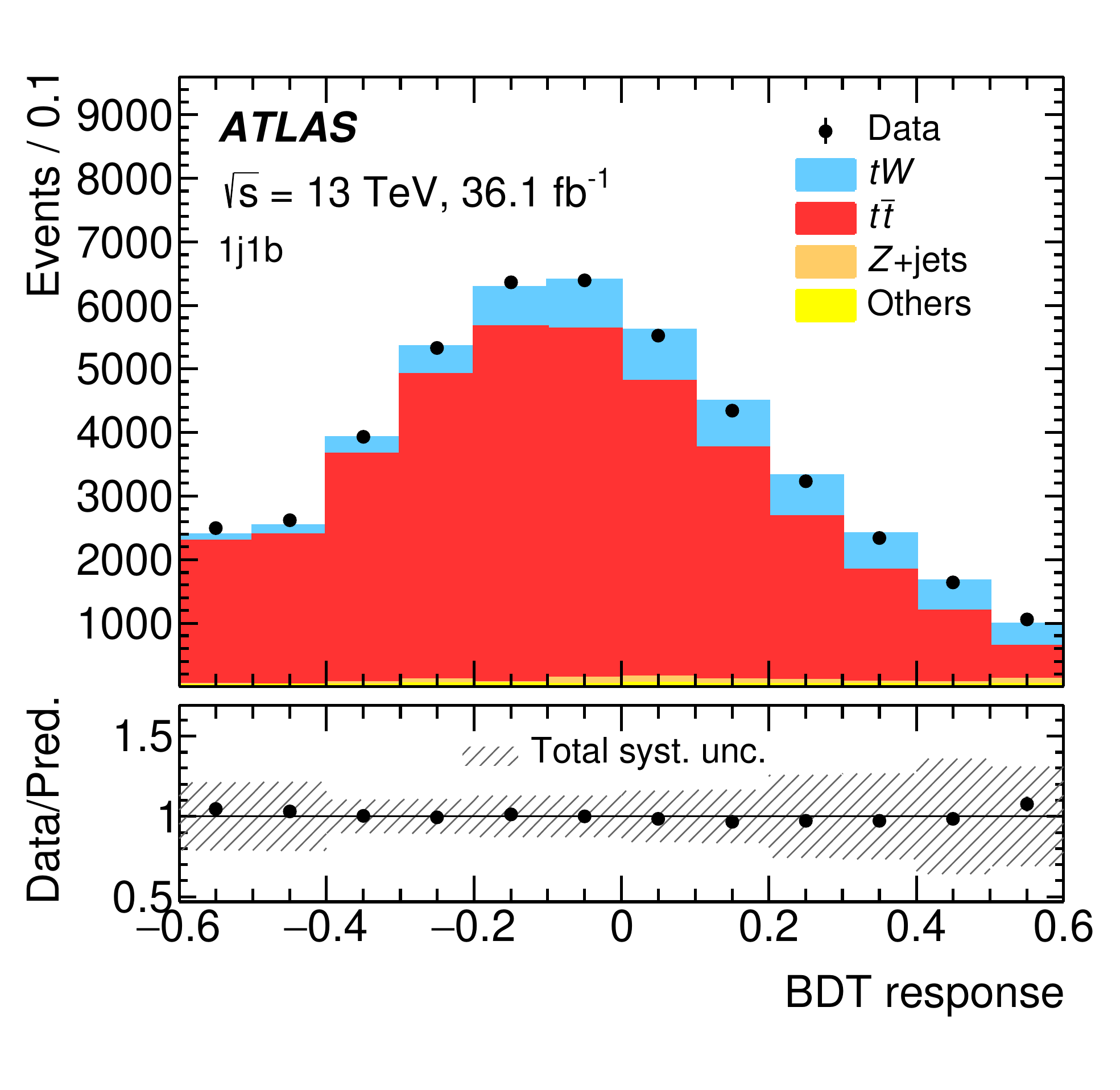}
\caption{\label{fig:plot_responseBDT_1j1b}
Comparison of data and MC predictions for the BDT response in the signal region.
The $tW$ signal is normalised with the measured fiducial cross-section.
Uncertainty bands reflect the total systematic uncertainties.
The first and last bins contain underflow and overflow events, respectively.}
\end{figure}

The final set of input variables used in the BDT is listed in \Tab{\ref{tab:mva_variables}} along with the separation power of each variable.\footnote{The separation power, $S$, is a measure of the difference between probability distributions of signal and background in the variable, and is defined as:
\begin{equation*}
\langle S^2 \rangle = \frac{1}{2} \int{
\frac{{\left(Y_{\text{s}}(y) - Y_{\text{b}}(y)\right)}^2}
{Y_{\text{s}}(y)+Y_{\text{b}}(y)} \text{d}y}
\end{equation*}
where $Y_{\text{s}}(y)$ and $Y_{\text{b}}(y)$ are the signal and background probability distribution functions of each variable $y$, respectively.}
The distributions of these variables are compared between the MC predictions and observed data, and found to be well modelled.
The BDT discriminant distributions from MC predictions and data are compared and shown in \Fig{\ref{fig:plot_responseBDT_1j1b}}.
 
\begin{table}
\caption{\label{tab:mva_variables}
The variables used in the signal region BDT and their separation power (denoted $S$).
The variables are derived from the four-momenta of the leading lepton ($\ell_1$), sub-leading lepton ($\ell_2$), the $b$-jet ($b$) and \met{}.
The last row gives the separation power of the BDT discriminant response.
}
\centering
\renewcommand{\arraystretch}{1.2}
\begin{tabular}{lr}
\toprule
Variable & \multicolumn{1}{c}{$S~$[$10^{-2}$]} \\
\midrule \
$\pt          (\ell_{1} \ell_{2} \MET b)   $ & 4.1 \\
$\Delta\pT    (\ell_{1} \ell_{2} b, \MET)  $ & 2.5 \\
$\sum{\ET}                                 $ & 2.3 \\
$\eta         (\ell_{1} \ell_{2} \MET b)   $ & 1.3 \\
$\Delta\pT    (\ell_{1} \ell_{2}, \MET)    $ & 1.1 \\
$\pt          (\ell_{1} \ell_{2} b)        $ & 1.0 \\
$C            (\ell_{1} \ell_{2})          $ & 0.9 \\
$m            (\ell_{2}, b)                $ & 0.2 \\
$m            (\ell_{1}, b)                $ & 0.1 \\
\midrule
BDT response                               &8.1\\
\bottomrule
\end{tabular}
\end{table}
 
To select a signal-enriched portion of events in the signal region, the BDT response is required to be larger than $0.3$.
The effect of this requirement on event yields is shown in \Tab{\ref{tab:yields}}.
The BDT requirement lowers systematic uncertainties by reducing contributions from the \ttbar{} background, which is subject to large modeling uncertainties.
For example, the total systematic uncertainty in the fiducial cross-section is reduced by $16\%$ of the total when applying the BDT response requirement, compared to having no requirement.
The exact value of the requirement is optimised to reduce the total uncertainty of the measurement over all bins, considering both statistical and systematic uncertainties.
 
\section{Unfolding and cross-section determination}
The iterative Bayesian unfolding technique in \Ref{\cite{DAgostini:1994zf}}, as implemented in the \texttt{RooUnfold} software package~\cite{Adye:2011gm}, is used to correct for detector acceptance and resolution effects and the efficiency to pass the event selection.
The unfolding procedure includes bin-by-bin correction for out-of-fiducial ($C_j^\text{oof}$) events which are reconstructed but fall outside the fiducial acceptance at particle level:
\begin{linenomath}
\begin{equation*}
C_j^{\text{oof}} = \frac{N^\text{fid\&reco}}{N^\text{reco}}.
\end{equation*}
\end{linenomath}
followed by the iterative matrix unfolding procedure.
The matrix $M$ is the migration matrix, and $M^{-1}$ represents the application of the iterative unfolding procedure with migration information from $M$.
The iterative unfolding is followed by another bin-by-bin correction to the efficiency to reconstruct a fiducial event ($C_i^{\text{eff}}$):
\begin{linenomath}
\begin{equation*}
\frac{1}{C_i^{\text{eff}}} = \frac{N^\text{fid}}{N^\text{fid\&reco}},
\end{equation*}
\end{linenomath}
In both expressions, ``fid'' refers to events passing the fiducial selection, ``reco'' refers to events passing reconstruction-level requirements, and ``fid\&reco'' refers to events passing both.
This full unfolding procedure is then described by the expression for the number of unfolded events in bin $i$ ($N_i^{\text{ufd}}$) of the particle-level distribution:
\begin{linenomath}
\begin{equation*}
N_i^{\text{ufd}} = \frac{1}{C_i^{\text{eff}}} \sum_j M_{ij}^{-1} C_j^{\text{oof}} (N_j^{\text{data}} - B_j),
\end{equation*}
\end{linenomath}
where $i$ ($j$) indicates the bin at particle (reconstruction) level, $N_j^{\text{data}}$ is the number of events in data and $B_j$ is the sum of all background contributions.
\Tab{\ref{tab:niter_summary}} gives the number of iterations used for each observable in this unfolding step.
The bias is defined as the difference between the unfolded and true values.
The number of iterations is chosen to minimise the growth of the statistical uncertainty propagated through the unfolding procedure while operating in a regime where the bias is sufficiently independent of the number of iterations.
The optimal number of iterations is small for most observables, but a larger number is picked for \EJet, where larger off-diagonal elements of the migration matrix cause slower convergence of the method.
 
The list of observables chosen was also checked for shaping induced by the requirement on the BDT response, since strong shaping can make the unfolding unstable.
These shaping effects were found to be consistently well-described by the various MC models considered.
Any residual differences in the predictions of different MC event generators would increase MC modelling uncertainties, thus ensuring shaping effects of the BDT are covered by the total uncertainties.
\begin{table}
\caption{\label{tab:niter_summary}
Number of iterations chosen in the unfolding procedure for each of the observables used in the measurement.}
\centering
\renewcommand{\arraystretch}{1.2}
\begin{tabular}{lc}
\toprule
Observable & Number of iterations \\
\midrule
\EJet{}                     & 15 \\
\massLepOneJet{}            & 7 \\
\massLepTwoJet{}            & 5 \\
\ELepLepJet{}               & 5 \\
\massTransLepLepMETJet{}    & 7 \\
\massLepLepJet{}            & 5 \\
\bottomrule
\end{tabular}
\end{table}
 
Unfolded event yields $N_i^\text{ufd}$ are converted to cross-section values as a function of an observable $X$ using the expression:
\begin{linenomath}
\begin{equation*}
\frac{\text{d}\sigma_i}{\text{d}X} =  \frac{N_i^{\text{ufd}}}{L \Delta_i},
\end{equation*}
\end{linenomath}
where $L$ is the integrated luminosity of the data sample and $\Delta_i$ is the width of bin $i$ of the particle-level distribution.
Differential cross-sections are divided by the fiducial cross-section to create a normalised distribution.
The fiducial cross-section is simply the sum of the cross-sections in each bin multiplied by the corresponding bin widths:
\begin{linenomath}
\begin{equation*}
\sigma^{\text{fid}} =  \sum_{i} \left( \frac{\text{d}\sigma_i}{\text{d}X} \cdot \Delta_i \right) = \sum_{i} \frac{N_i^{\text{ufd}}}{L}.
\end{equation*}
\end{linenomath}
 
\section{Systematic uncertainties}
\label{sec:systematics}
\subsection{Sources of systematic uncertainty}
The experimental sources of uncertainty include the uncertainty in the lepton efficiency scale factors used to correct simulation to data, the lepton energy scale and resolution, the \met{} soft-term calculation, the jet energy scale and resolution, the \btag{} efficiency, and the luminosity.
 
The JES uncertainty~\cite{ATL-PHYS-PUB-2015-015} is divided into 18 components, which are derived using $\sqrt{s}=\SI{13}{\TeV}$ data.
The uncertainties from data-driven calibration studies of $Z/\gamma +$jet and dijet events are represented with six orthogonal components using the eigenvector decomposition procedure, as demonstrated in Ref.~\cite{PERF-2012-01}.
Other components include model uncertainties (such as flavour composition, $\eta$ intercalibration model).
The most significant JES uncertainty components for this measurement are the data-driven calibration and the flavour composition uncertainty, which is the dependence of the jet calibration on the fraction of quark or gluon jets in data.
The jet energy resolution uncertainty estimate~\cite{ATL-PHYS-PUB-2015-015} is based on comparisons of simulation and data using studies of Run-1 data.
These studies are then cross-calibrated and checked to confirm good agreement with Run-2 data.
 
As discussed in \Sect{\ref{sec:objects}}, the \met{} calculation includes contributions from leptons and jets in addition to soft terms which arise primarily from low-\pt{} pile-up jets and underlying-event activity~\cite{ATL-PHYS-PUB-2015-023,ATL-PHYS-PUB-2015-027}.
The uncertainty associated with the leptons and jets is propagated from the corresponding uncertainties in the energy/momentum scales and resolutions, and it is classified together with the uncertainty associated with the corresponding objects.
The uncertainty associated with the soft term is estimated by comparing the simulated soft-jet energy scale and resolution to that in data.
 
Uncertainties in the scale factors used to correct the \btag{} efficiency in simulation to the efficiency in data are assessed using independent eigenvectors for the efficiency of $b$-jets, $c$-jets, light-parton jets, and the extrapolation uncertainty for high-\pt{} jets~\cite{PERF-2012-04,ATL-PHYS-PUB-2016-012}.
 
Systematic uncertainties in lepton momentum resolution and scale, trigger efficiency, isolation efficiency, and identification efficiency are also considered~\cite{ATLAS-CONF-2016-024,PERF-2015-10,PERF-2014-07}.
These uncertainties arise from corrections to simulation based on studies of $Z \to ee$ and $Z \to \mu\mu$ data.
In this measurement, the effects of the uncertainties in these corrections are relatively small.
 
A \lumiUncPct{\%} uncertainty is assigned to the integrated luminosity.
It is derived, following a methodology similar to that detailed in \Ref{\cite{DAPR-2013-01}}, from a calibration of the luminosity scale using $x$--$y$ beam-separation scans.
 
Uncertainties stemming from theoretical models are estimated by comparing a set of predicted distributions produced with different assumptions.
The main uncertainties are due to the NLO matrix-element (ME) event generator, parton shower and hadronisation event generator, radiation tuning and scale choice and the PDF\@.
The NLO matrix-element uncertainty is estimated by comparing two NLO matching methods: the predictions of \POWHEGBOX{} and \MGMCatNLO, both interfaced with \HERWIGpp.
The parton shower, hadronisation, and underlying-event model uncertainty is estimated by comparing \POWHEGBOX{} interfaced with either \PYTHIA6 or \textsc{Herwig++}.
The uncertainty from the matrix-element event generator is treated as uncorrelated between the $tW$ and \ttbar{} processes, while the uncertainty from the parton shower event generator is treated as correlated.
The radiation tuning and scale choice uncertainty is estimated by taking half of the difference between samples with \POWHEGBOX{} interfaced with \PYTHIAV{6} tuned with either more or less radiation, and is uncorrelated between the $tW$ and \ttbar{} processes.
These choices of correlations are based on \Ref{\cite{TOPQ-2015-16}}, and were checked to be no less conservative than the alternative options.
The choice of scheme to account for the interference between the $tW$ and \ttbar{} processes constitutes another source of systematic uncertainty for the signal modelling, and it is estimated by comparing samples using either the diagram removal scheme or the diagram subtraction scheme, both generated with \POWHEGBOX{}+\PYTHIAV{6}.
The uncertainty due to the choice of PDF is estimated using the PDF4LHC15 combined PDF set~\cite{Butterworth:2015oua}.
The difference between the central CT10~\cite{ct10} prediction and the central PDF4LHC15 prediction (PDF central value) is taken and symmetrised together with the internal uncertainty set provided with PDF4LHC15.
 
Additional normalisation uncertainties are applied to each background.
A $100\%$ uncertainty is applied to the normalisation of the background from non-prompt and fake leptons, an uncertainty of $50\%$ is applied to the \Zjets{} background, and a $25\%$ normalisation uncertainty is assigned to diboson backgrounds.
These uncertainties are based on earlier ATLAS studies of background simulation in top quark analyses~\cite{TOPQ-2015-09}.
These normalisation uncertainties are not found to have a large impact on the final measurement due to the small contribution of these backgrounds in the signal region as well as their cancellation in the normalised cross-section measurement.
An uncertainty of $5.5\%$ is applied to the \ttbar{} normalisation to account for the scale, $\alphas$, and PDF uncertainties in the NNLO cross-section calculation.
 
Uncertainties due to the size of the MC samples are estimated using pseudoexperiments.
An ensemble of pseudodata is created by fluctuating the MC samples within the statistical uncertainties.
Each set of pseudodata is used to construct $M_{ij}$, $C_i^\text{eff}$, and $C_j^\text{oof}$, and the nominal MC sample is unfolded.
The width of the distribution of unfolded values from this ensemble is taken as the statistical uncertainty.
Additional non-closure uncertainties are added in certain cases after stress-testing the unfolding procedure with injected Gaussian or linear functions.
Each distribution is tested by reweighting the input MC sample according to the injected function, unfolding, and checking that the weights are recovered in the unfolded distribution.
The extent to which the unfolded weighted data are biased with respect to the underlying weighted generator-level distribution is taken as the unfolding non-closure uncertainty.
 
\subsection{Procedure for estimation of uncertainty}
The propagation of uncertainties through the unfolding process proceeds by constructing the migration matrix and efficiency corrections with the baseline sample and unfolding with the varied sample as input.
In most cases, the baseline sample is from \POWHEGBOX{}+\PYTHIAV{6} and produced with the full detector simulation, but in cases where the varied sample uses the \textsc{Atlfast2} fast simulation, the baseline sample is also changed to use \textsc{Atlfast2}.
For uncertainties modifying background processes, varied samples are prepared by taking into account the changes in the background induced by a particular systematic effect.
Experimental uncertainties are treated as correlated between signal and background in this procedure.
The varied samples are unfolded and compared to the corresponding particle-level distribution from the MC event generator; the relative difference in each bin is the estimated systematic uncertainty.
 
The covariance matrix $\mathbf{C}$ for each differential cross-section measurement is computed following a procedure similar to that used in \Ref{\cite{TOPQ-2014-15}}.
Two covariance matrices are summed to form the final covariance.
The first one is computed using $\num{10000}$ pseudoexperiments and includes statistical uncertainties as well as systematic uncertainties from experimental sources.
The statistical uncertainties are included by independently fluctuating each bin of the data distribution according to Poisson distributions for each pseudoexperiment.
Each bin of the resulting pseudodata distribution is then fluctuated according to a Gaussian distribution for each experimental uncertainty, preserving bin-to-bin correlation information for each uncertainty.
The other matrix includes the systematic uncertainties from event generator model uncertainties, PDF uncertainties, unfolding non-closure uncertainties, and MC statistical uncertainties.
In this second matrix, the bin-to-bin correlation value is set to zero for the non-closure and MC statistical uncertainties, and set to unity for the other uncertainties.
The impact of setting the bin-to-bin correlation value to unity was compared for the non-closure uncertainty, and this choice was found to have negligible impact on the results.
This covariance matrix is used to compute a $\chi^2$ and corresponding $p$-value to assess how well the measurements agree with the predictions.
The $\chi^2$ values are computed using the expression:
\begin{linenomath}
\begin{equation*}
\chi^2 = \mathbf{v}^\intercal \mathbf{C}^{-1} \mathbf{v},
\end{equation*}
\end{linenomath}
where $\mathbf{v}$ is the vector of differences between the measured cross-sections and predictions.
 
\section{Results}
\begin{figure}[htbp]
\begin{tabular}{cc}
\includegraphics[width=.45\textwidth]{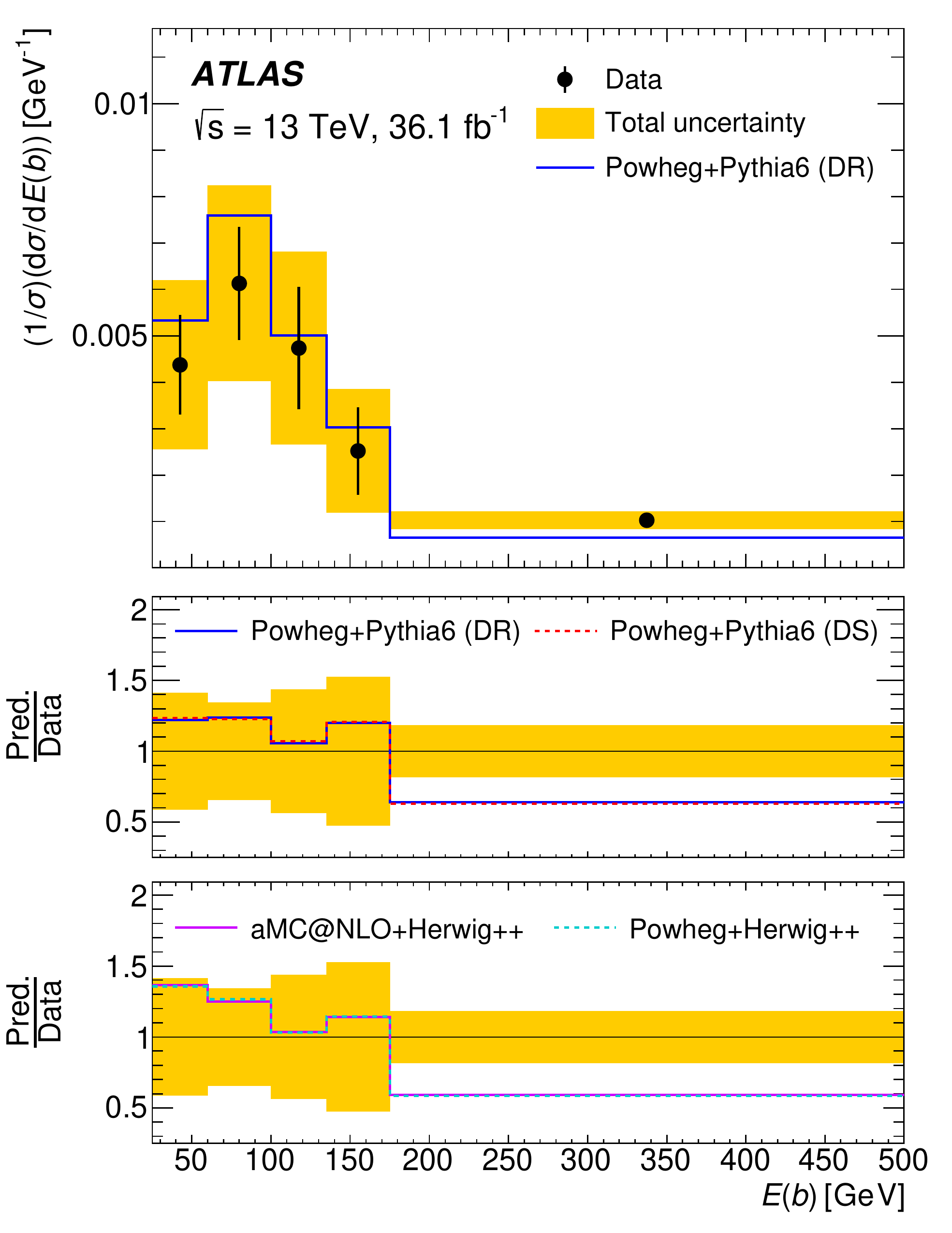} &
\includegraphics[width=.45\textwidth]{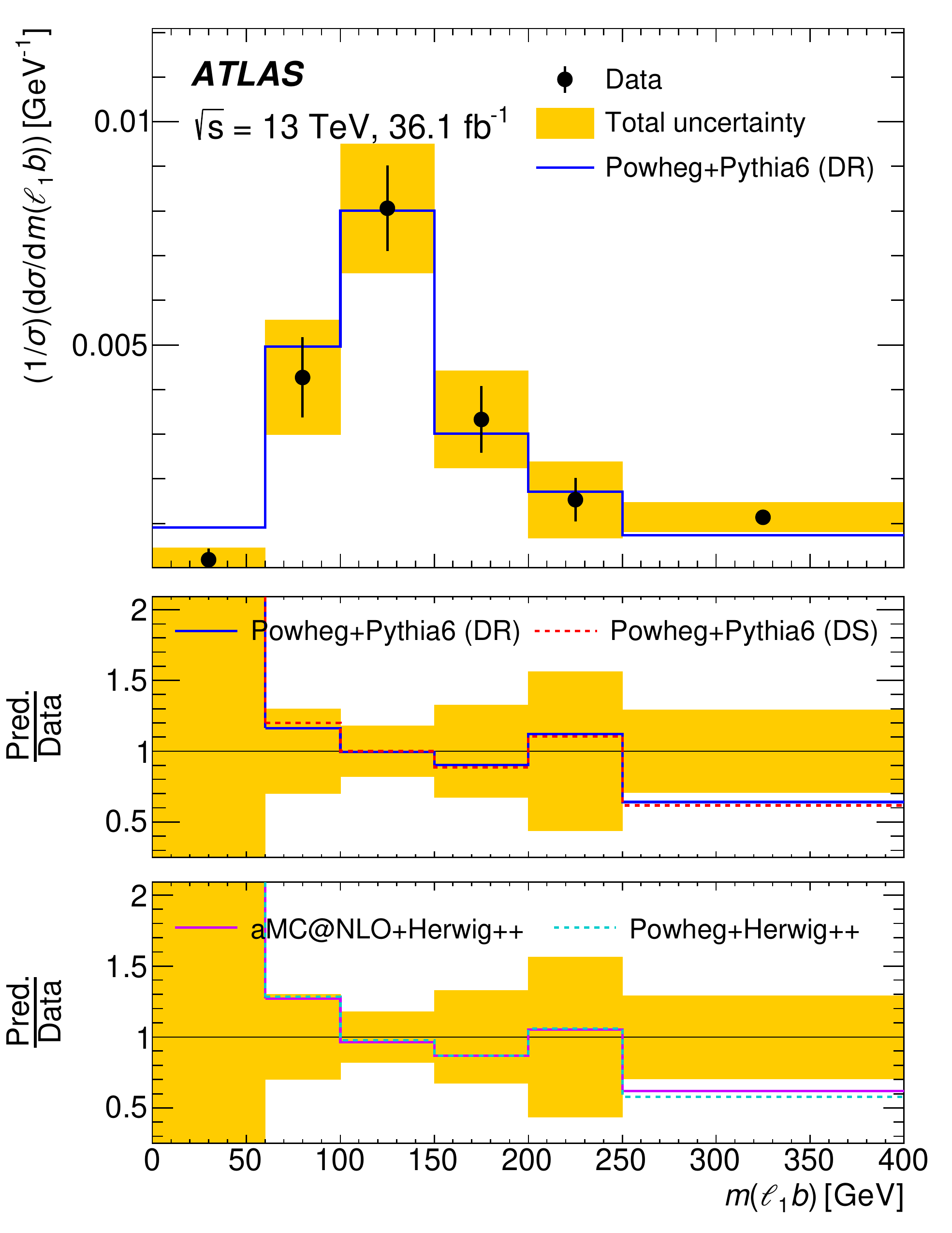} \\
\includegraphics[width=.45\textwidth]{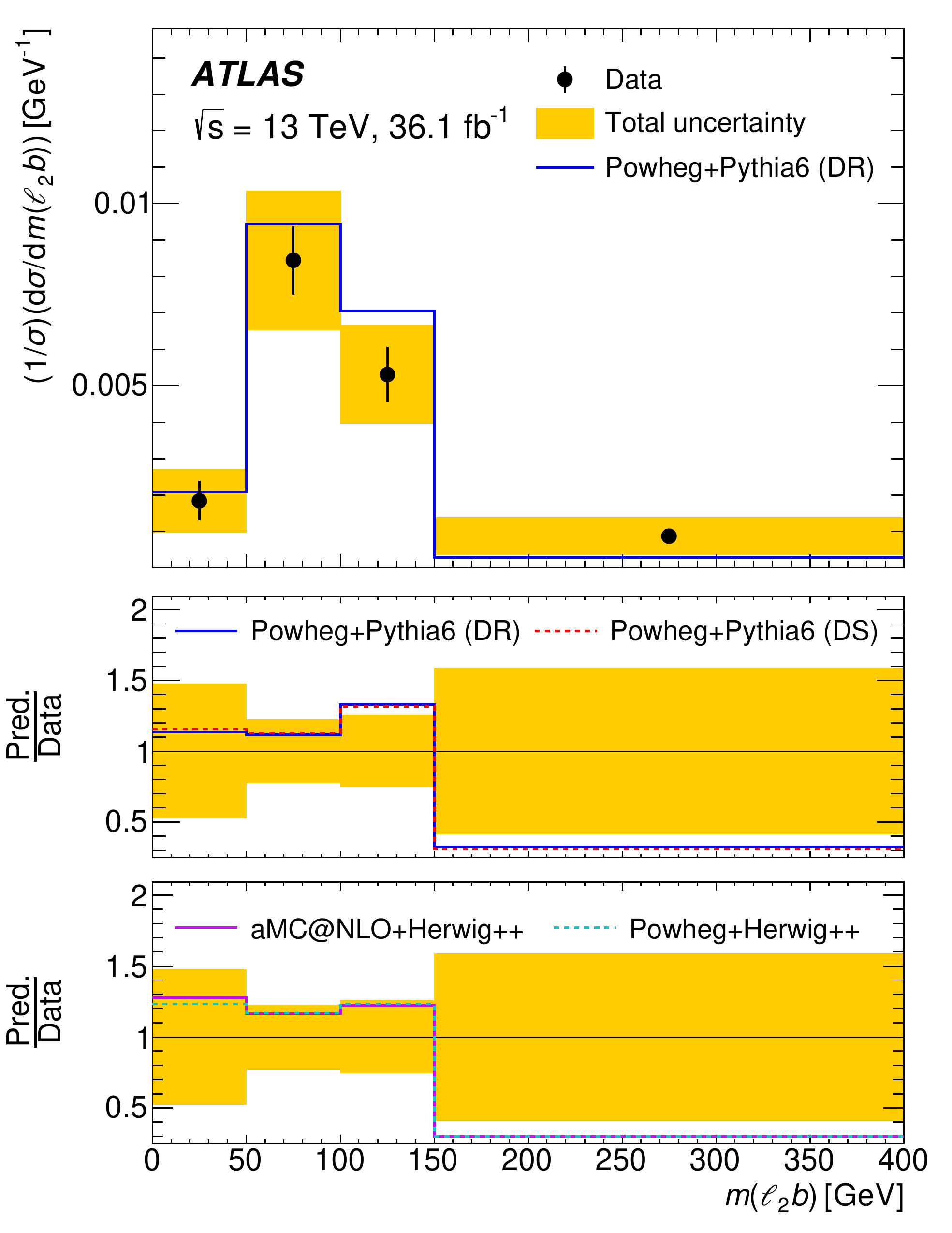} &
\includegraphics[width=.45\textwidth]{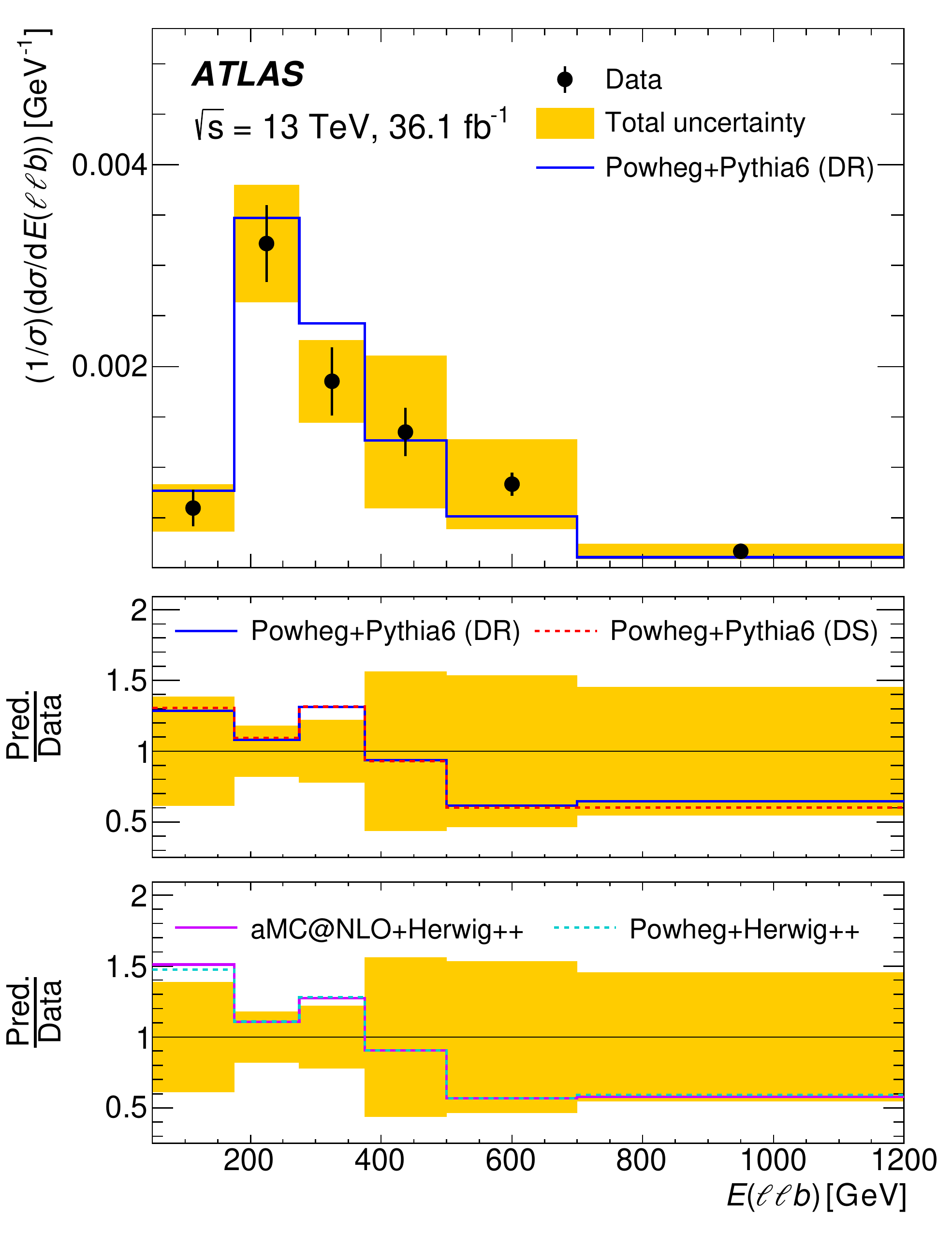} \\
\end{tabular}
\caption{\label{fig:result_norm1}
Normalised differential cross-sections unfolded from data, compared with selected MC models, with respect to \EJet, \massLepOneJet, \massLepTwoJet, and \ELepLepJet.
Data points are placed at the horizontal centre of each bin, and the error bars on the data points show the statistical uncertainties.
The total uncertainty in the first bin of the \massLepOneJet{} distribution (not shown) is $140\%$.
See \Sect{\ref{sec:intro}} for a description of the observables plotted.}
\end{figure}

\begin{figure}[htbp]
\begin{tabular}{cc}
\includegraphics[width=.45\textwidth]{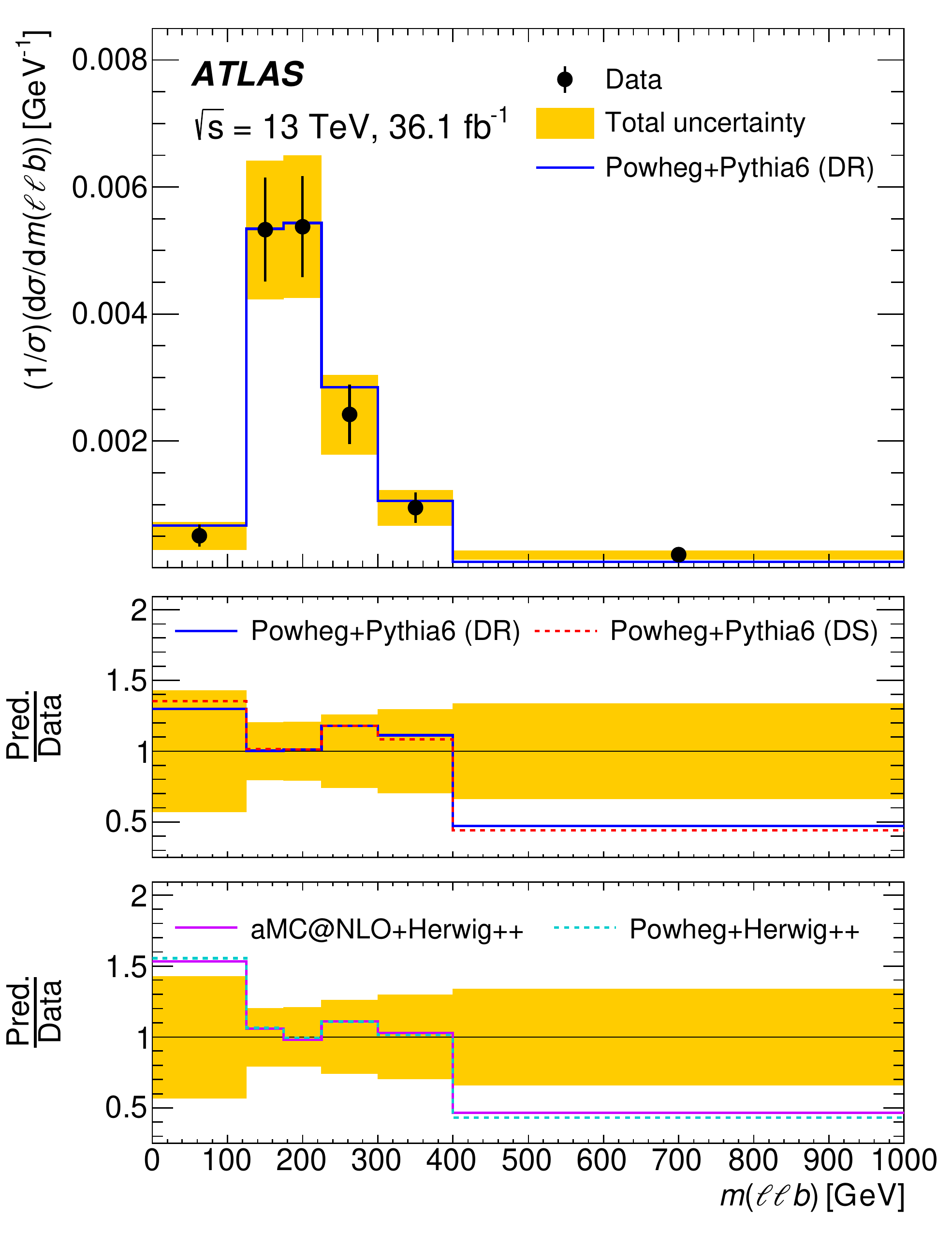} &
\includegraphics[width=.45\textwidth]{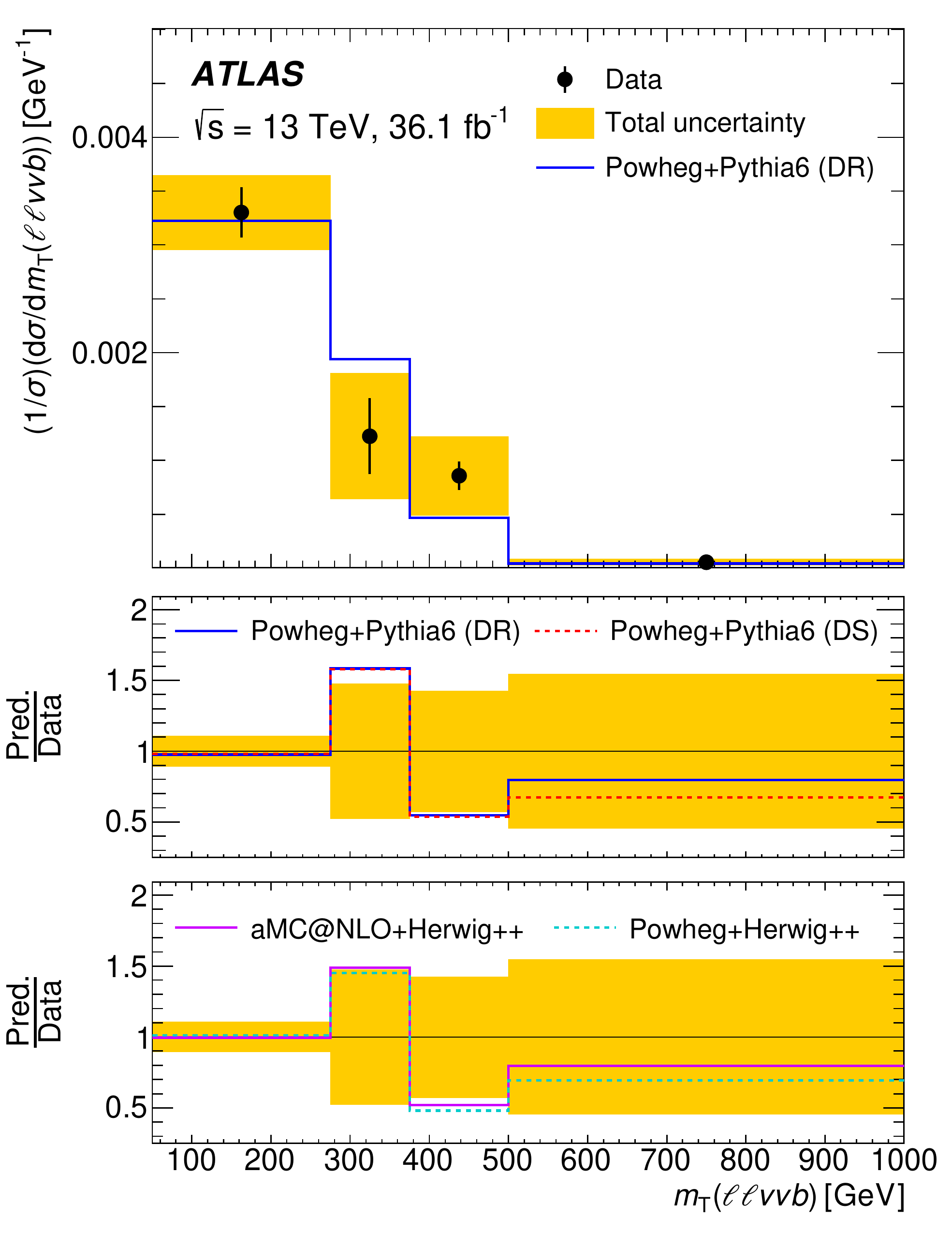} \\
\end{tabular}
\caption{\label{fig:result_norm2}
Normalised differential cross-sections unfolded from data, compared with selected MC models, with respect to \massTransLepLepMETJet{} and \massLepLepJet.
Data points are placed at the horizontal centre of each bin.
See \Sect{\ref{sec:intro}} for a description of the observables plotted.}
\end{figure}

\begin{figure}[htbp]
\begin{tabular}{cc}
\includegraphics[width=.45\textwidth]{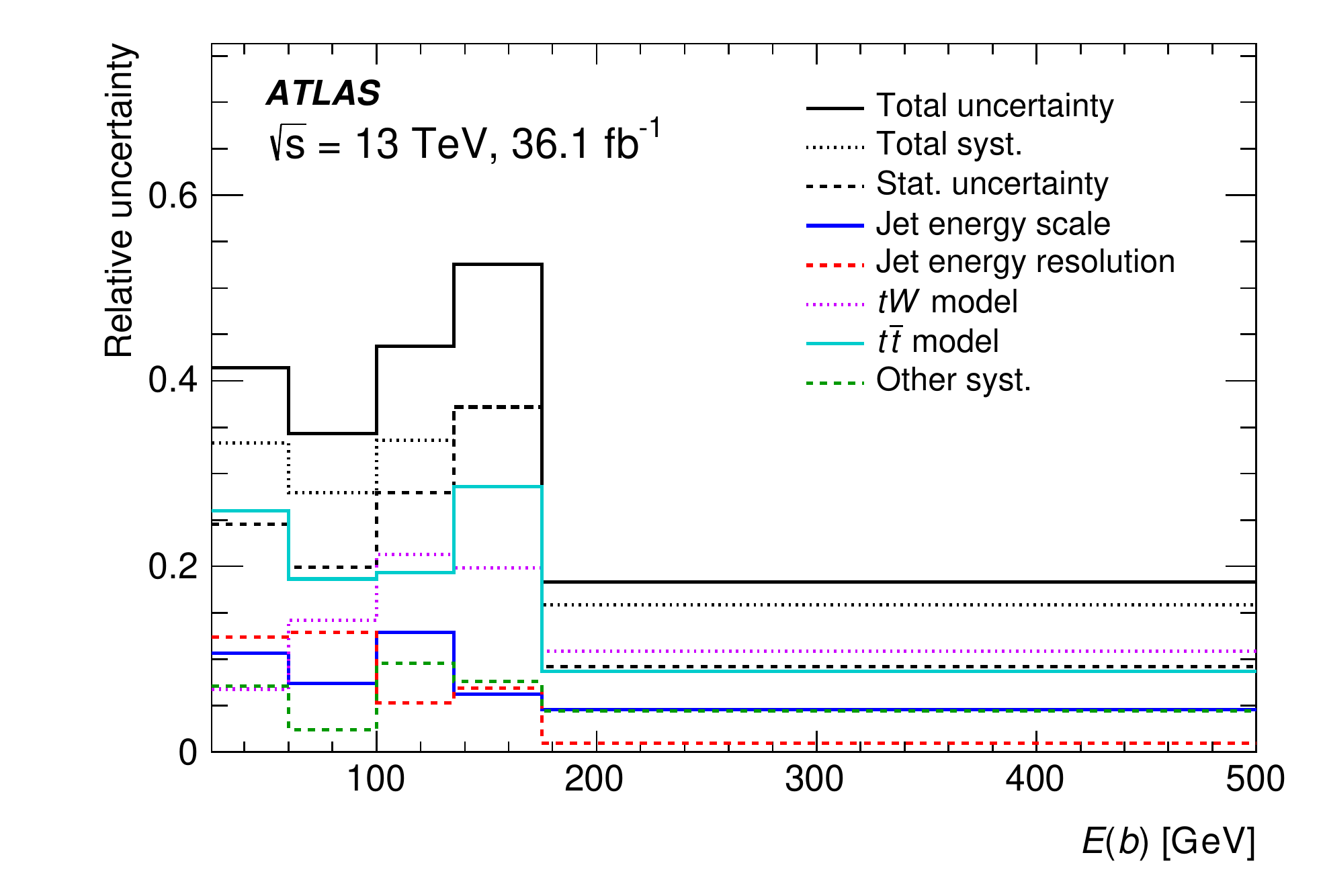} &
\includegraphics[width=.45\textwidth]{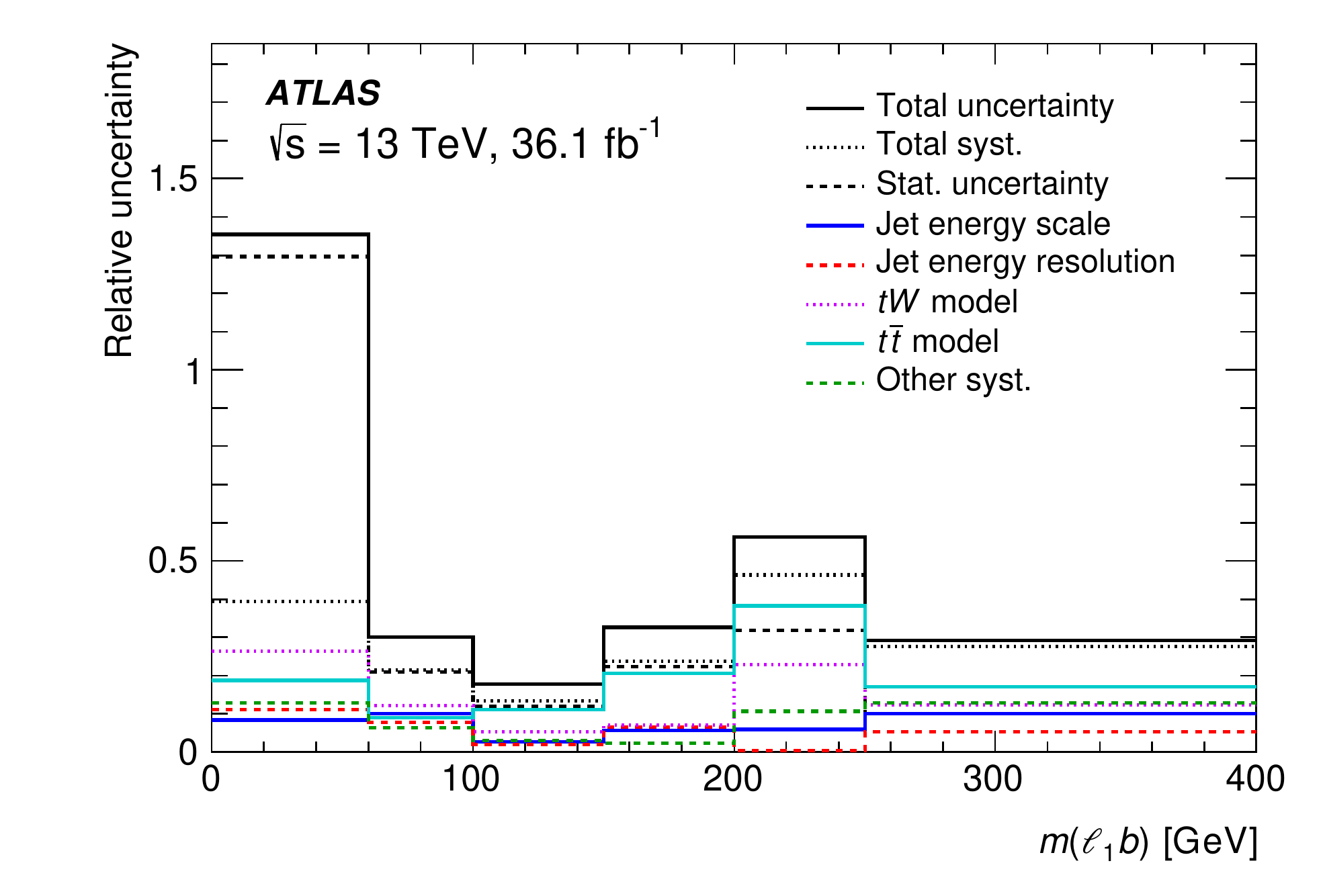} \\
\includegraphics[width=.45\textwidth]{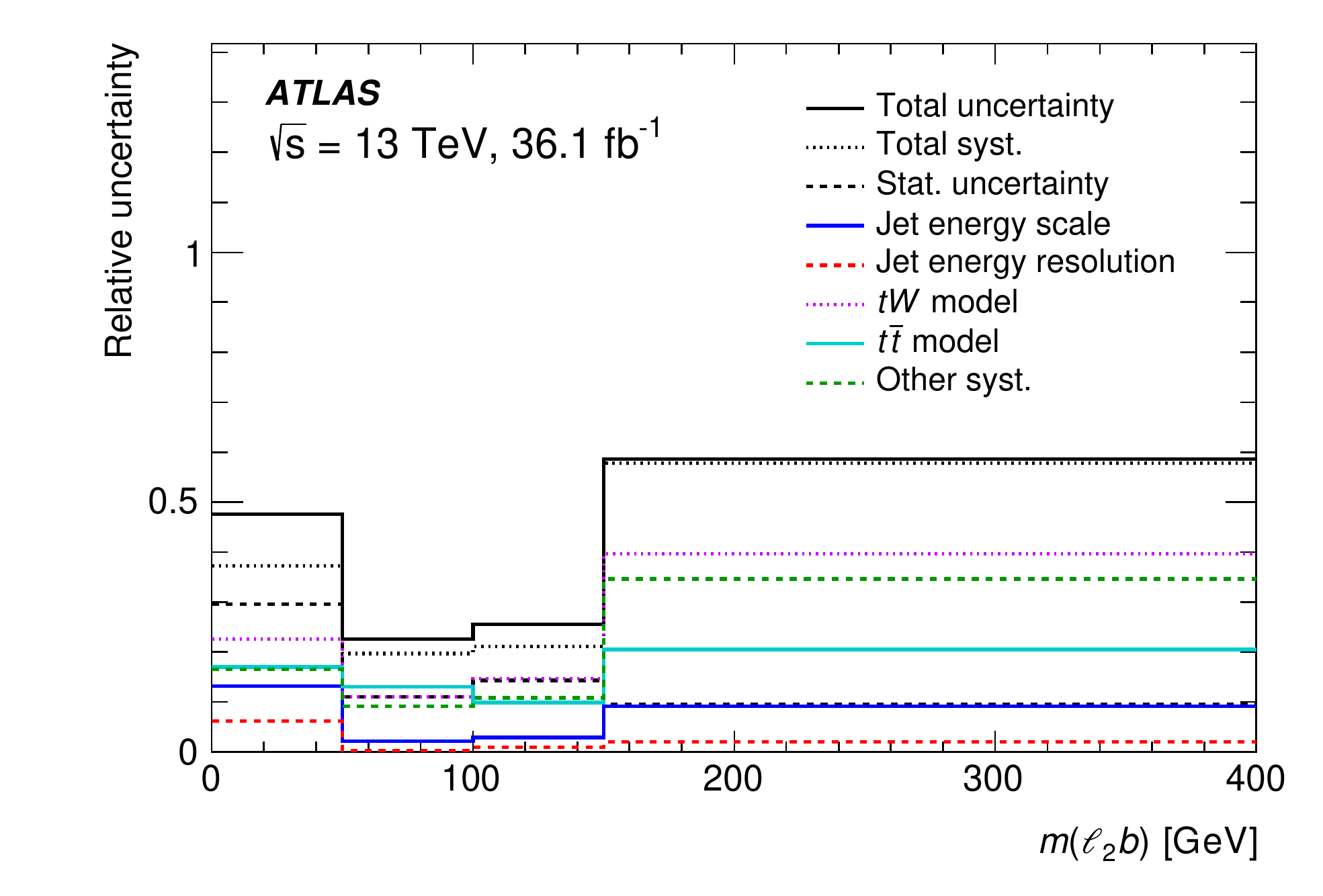} &
\includegraphics[width=.45\textwidth]{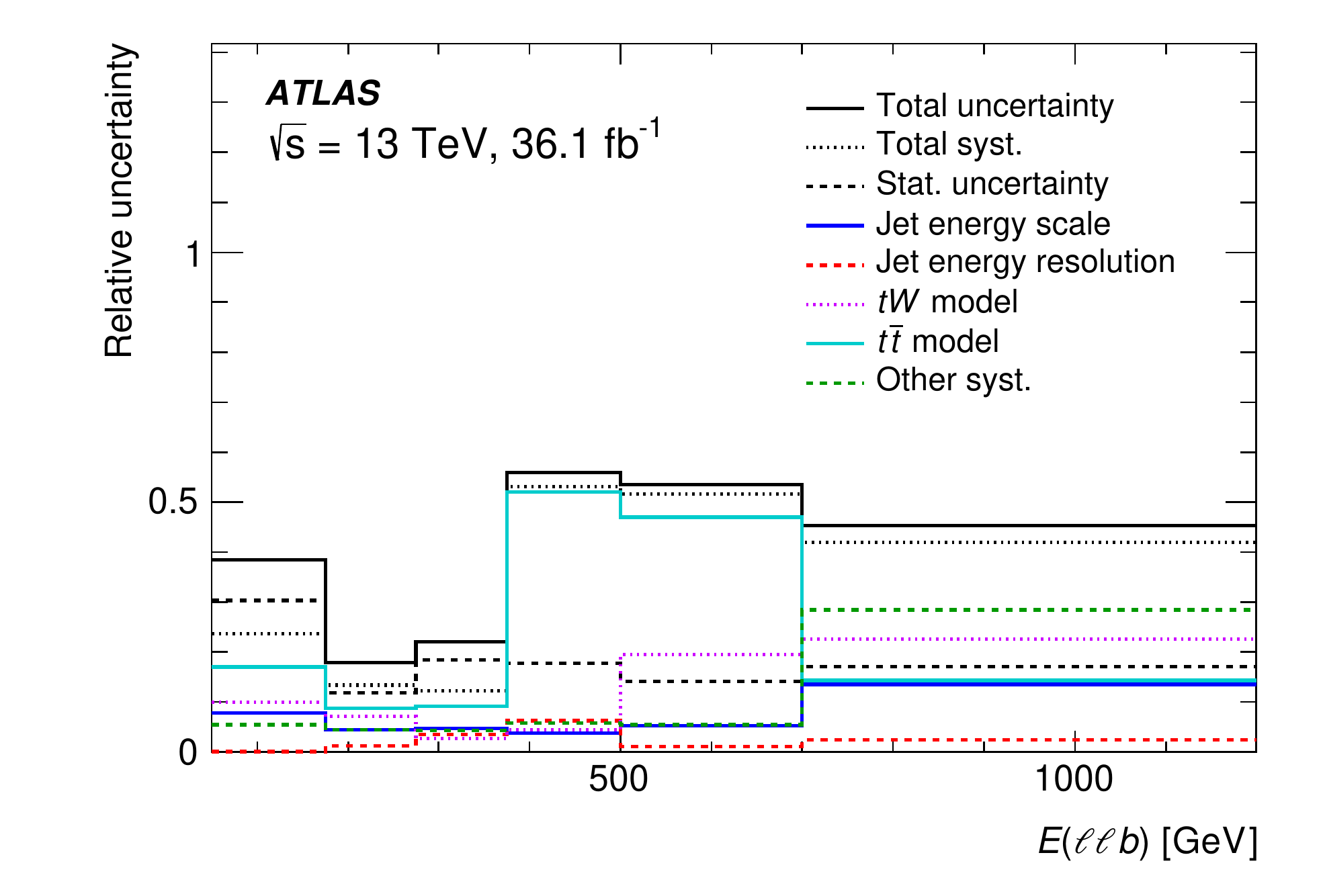} \\
\includegraphics[width=.45\textwidth]{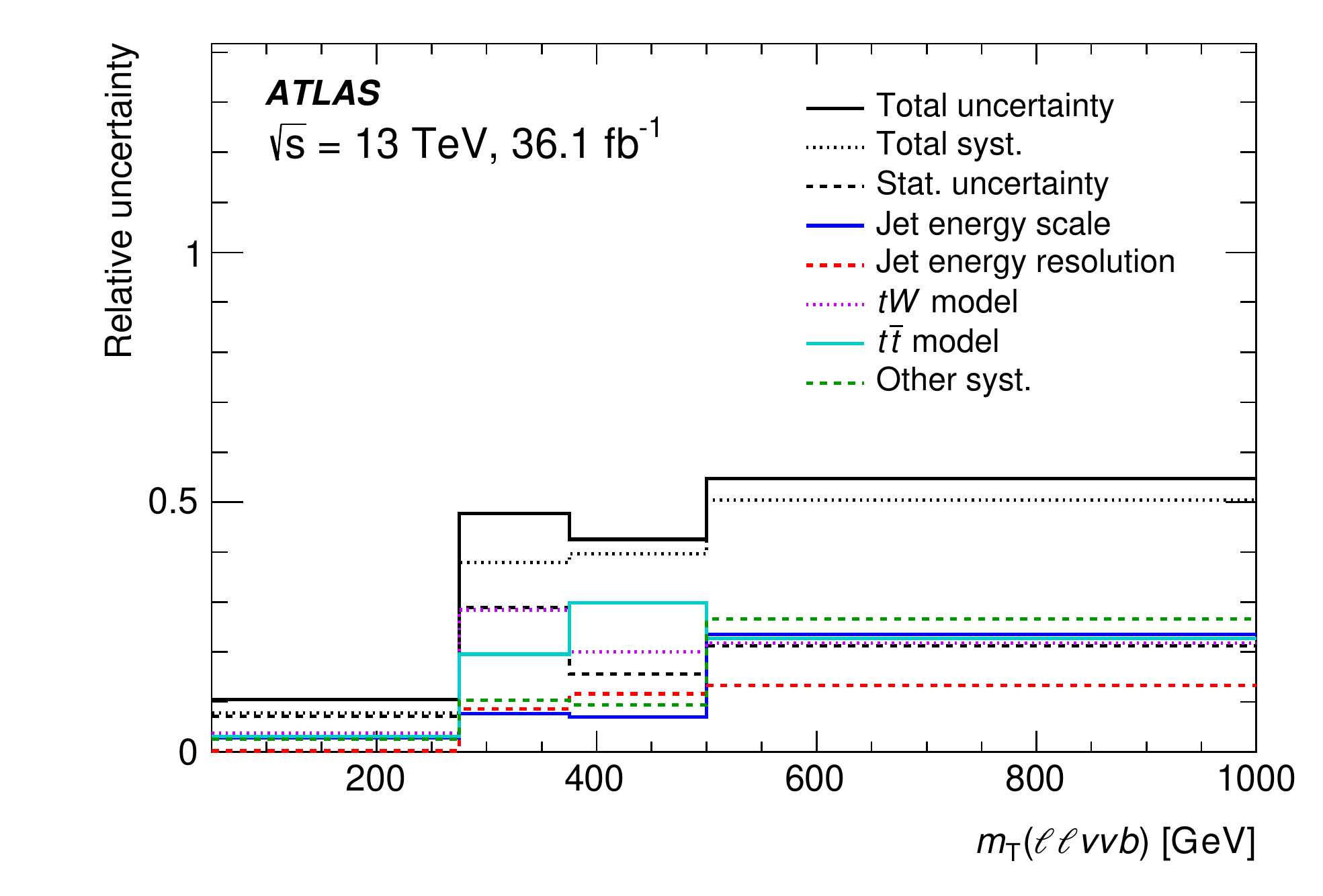} &
\includegraphics[width=.45\textwidth]{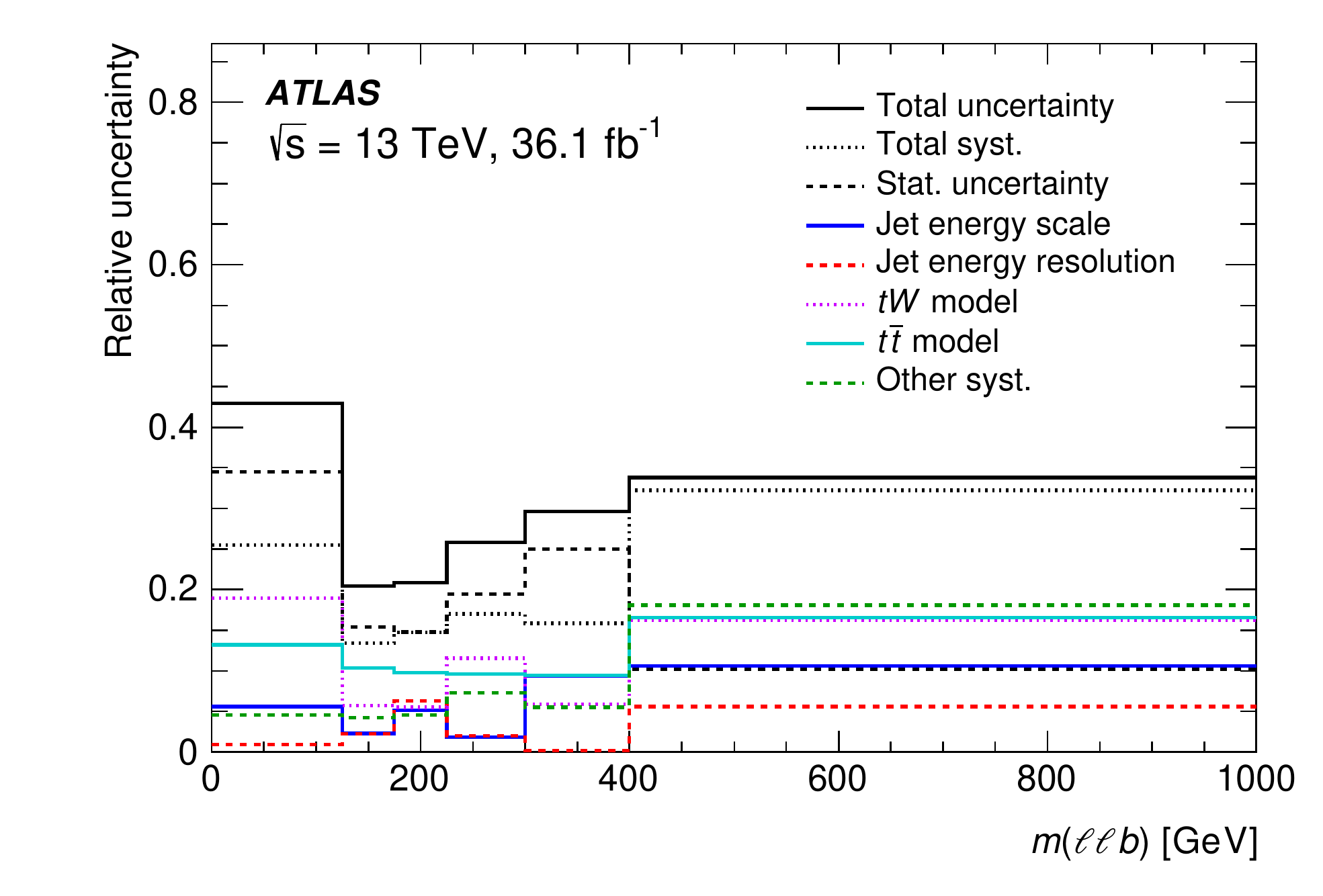} \\
\end{tabular}
\caption{\label{fig:result_syst_norm}
Summary of uncertainties in normalised differential cross-sections unfolded from data.}
\end{figure}

Unfolded particle-level normalised differential cross-sections are given in \Tab{\ref{tab:result__WtOnly__Normalised}}.
In \Figrange{\ref{fig:result_norm1}}{\ref{fig:result_norm2}}, the results are shown compared to the predictions of various MC event generators, and in \Fig{\ref{fig:result_syst_norm}} the main systematic uncertainties for each distribution are summarised.
The results show that the largest uncertainties come from the size of the data sample as well as \ttbar{} and $tW$ MC modelling.

The comparison between the data and Monte Carlo predictions is summarised in
\Tab{\ref{tab:chi2-1}}, where $\chi^2$ values and
corresponding $p$-values are listed.
In general, most of the MC models show fair agreement with the measured
cross-sections, with no particularly low $p$-values observed.  Notably, for each
distribution there is a substantial negative slope in the ratio of predicted to
observed cross-sections, indicating there are more events with high-momentum
final-state objects than several of the MC models predict.  This effect is most
visible in the \ELepLepJet{} distribution, where the lower $p$-values for all MC
predictions reflect this.  In most cases, differences between the MC predictions
are smaller than the uncertainty on the data, but there
are some signs that \POWHEGBOX{}+\HERWIGpp{} deviates more from the data and from
the other predictions in certain bins of the \ELepLepJet{}, \massLepLepJet{}, and
\massLepOneJet{} distributions. The predictions of DS and DR samples likewise give
very similar results for all observables as expected from the fiducial selection.
The predictions of \POWHEGBOX{}+\PYTHIAV{6}
with varied initial- and final-state radiation tuning were also examined but not
found to give significantly different distributions in the fiducial phase space
of this analysis.
 
Both the statistical and systematic uncertainties have a significant impact on the
result. The exact composition varies bin-to-bin but there is no single source
of uncertainty that dominates each normalised measurement.
Some of the largest systematic uncertainties are those related to \ttbar{} and $tW$ modelling.
The cancellation in the normalised differential cross-sections is very
effective at reducing a number of systematic uncertainties. The most notable
cancellation is related to the \ttbar{} parton shower model uncertainty, which is
quite dominant prior to dividing by the fiducial cross-section.
 
\begin{DIFnomarkup}
\begin{table}[htbp]
\renewcommand{\arraystretch}{1.1}
\caption{\label{tab:result__WtOnly__Normalised}
Summary of the measured normalised differential cross-sections, with uncertainties shown as percentages.
The uncertainties are divided into statistical and systematic contributions.
}
\centering
\begin{tabular}{lrrrrrr}
\toprule
\EJet{} bin [GeV] & [25, 60] & [60, 100] & [100, 135] & [135, 175] & [175, 500] &\\
$(1/\sigma) \;\textrm{d}\sigma/\textrm{d}x$ [$\GeV^{-1}$] & 0.00438 & 0.00613 & 0.00474 & 0.00252 & 0.00103 &\\
\cmidrule{2-6}
Stat.\ uncertainty & 25 & 20 & 28 & 37 & 9.3 &\\
Total syst.\ uncertainty & 33 & 28 & 34 & 37 & 16 &\\
Total uncertainty & 41 & 34 & 44 & 53 & 18 &\\
\midrule
\massLepOneJet{} bin [GeV] & [0, 60] & [60, 100] & [100, 150] & [150, 200] & [200, 250] & [250, 400] \\
$(1/\sigma) \;\textrm{d}\sigma/\textrm{d}x$ [$\GeV^{-1}$] & 0.000191 & 0.00428 & 0.00806 & 0.00333 & 0.00153 & 0.00114 \\
\cmidrule{2-7}
Stat.\ uncertainty & 130 & 21 & 12 & 22 & 32 & 10 \\
Total syst.\ uncertainty & 39 & 22 & 13 & 24 & 46 & 28 \\
Total uncertainty & 140 & 30 & 18 & 33 & 56 & 29 \\
\midrule
\massLepTwoJet{} bin [GeV] & [0, 50] & [50, 100] & [100, 150] & [150, 400] &&\\
$(1/\sigma) \;\textrm{d}\sigma/\textrm{d}x$ [$\GeV^{-1}$] & 0.00184 & 0.00845 & 0.00531 & 0.000879 &&\\
\cmidrule{2-5}
Stat.\ uncertainty & 30 & 11 & 14 & 9.6 &&\\
Total syst.\ uncertainty & 37 & 20 & 21 & 58 &&\\
Total uncertainty & 48 & 23 & 25 & 59 &&\\
\midrule
\ELepLepJet{} bin [GeV] & [50, 175] & [175, 275] & [275, 375] & [375, 500] & [500, 700] & [700, 1200] \\
$(1/\sigma) \;\textrm{d}\sigma/\textrm{d}x$ [$\GeV^{-1}$] & 0.000597 & 0.00322 & 0.00185 & 0.00135 & 0.000832 & 0.000167 \\
\cmidrule{2-7}
Stat.\ uncertainty & 30 & 12 & 18 & 18 & 14 & 17 \\
Total syst.\ uncertainty & 24 & 13 & 12 & 53 & 52 & 42 \\
Total uncertainty & 38 & 18 & 22 & 56 & 53 & 45 \\
\midrule
\massTransLepLepMETJet{} bin [GeV] & [50, 275] & [275, 375] & [375, 500] & [500, 1000] &&\\
$(1/\sigma) \;\textrm{d}\sigma/\textrm{d}x$ [$\GeV^{-1}$] & 0.0033 & 0.00123 & 0.000856 & $5.51\times 10^{-5}$ &&\\
\cmidrule{2-5}
Stat.\ uncertainty & 7.1 & 29 & 16 & 21 &&\\
Total syst.\ uncertainty & 7.8 & 38 & 40 & 50 &&\\
Total uncertainty & 11 & 48 & 43 & 55 &&\\
\midrule
\massLepLepJet{} bin [GeV] & [0, 125] & [125, 175] & [175, 225] & [225, 300] & [300, 400] & [400, 1000] \\
$(1/\sigma) \;\textrm{d}\sigma/\textrm{d}x$ [$\GeV^{-1}$] & 0.00051 & 0.00533 & 0.00538 & 0.00242 & 0.000949 & 0.000208 \\
\cmidrule{2-7}
Stat.\ uncertainty & 35 & 15 & 15 & 19 & 25 & 10 \\
Total syst.\ uncertainty & 25 & 13 & 15 & 17 & 16 & 32 \\
Total uncertainty & 43 & 20 & 21 & 26 & 30 & 34 \\
\bottomrule
\end{tabular}
\end{table}
\end{DIFnomarkup}

\begin{DIFnomarkup}
\begin{table}[htbp]
\caption{\label{tab:chi2-1}
Values of $\chi^2$ and $p$-values for the measured normalised cross-sections compared to particle-level MC predictions.
}
\centering
\renewcommand{\arraystretch}{1.2}
\begin{tabular}{lllrlllllllll}
\toprule
Observable
& \multicolumn{2}{c}{\EJet}
& \multicolumn{2}{c}{\massLepOneJet}
& \multicolumn{2}{c}{\massLepTwoJet}
& \multicolumn{2}{c}{\ELepLepJet}
& \multicolumn{2}{c}{\massTransLepLepMETJet}
& \multicolumn{2}{c}{\massLepLepJet}
\\
Degrees of freedom
& \multicolumn{2}{c}{4}
& \multicolumn{2}{c}{5}
& \multicolumn{2}{c}{3}
& \multicolumn{2}{c}{5}
& \multicolumn{2}{c}{3}
& \multicolumn{2}{c}{5}
\\
Prediction
& $\chi^2$  & $p$
& $\chi^2$  & $p$
& $\chi^2$  & $p$
& $\chi^2$  & $p$
& $\chi^2$  & $p$
& $\chi^2$  & $p$
\\
\midrule
\POWHEG{}+\PYTHIAV{6} (DR)
& 4.8  & 0.31 &  5.7 & 0.34 &  2.6 & 0.45 & 8.1  & 0.15 & 2.0  & 0.56 &  4.0 & 0.55 \\
\POWHEG{}+\PYTHIAV{6} (DS)
& 5.0  & 0.29 &  6.1 & 0.30 &  2.6 & 0.46 & 9.1  & 0.11 & 2.4  & 0.49 &  4.4 & 0.50 \\
{aMC@NLO}+\HERWIGpp{}
& 5.6  & 0.23 &  5.4 & 0.37 &  2.4 & 0.49 & 8.7  & 0.12 & 1.8  & 0.61 &  3.6 & 0.61 \\
\POWHEG{}+\HERWIGpp{}
& 6.2  & 0.18 &  8.1 & 0.15 &  2.3 & 0.52 & 11.0 & 0.05 & 2.0  & 0.57 &  5.2 & 0.40 \\
\POWHEG{}+\PYTHIAV{6} radHi
& 4.8  & 0.30 &  5.3 & 0.38 &  2.5 & 0.48 & 7.9  & 0.16 & 1.9  & 0.60 &  3.7 & 0.60 \\
\POWHEG{}+\PYTHIAV{6} radLo
& 5.0  & 0.29 &  5.8 & 0.33 &  2.6 & 0.45 & 8.4  & 0.14 & 2.1  & 0.56 &  4.0 & 0.55 \\
\bottomrule
\end{tabular}
\end{table}
\end{DIFnomarkup}
 
\FloatBarrier{}
\section{Conclusion}
The differential cross-section for the production of a $W$ boson in association with a top quark is measured for several particle-level observables.
The measurements are performed using \lumi{} of $pp$ collision data with $\sqs=\SI{13}{TeV}$ collected in 2015 and 2016 by the ATLAS detector at the LHC\@.
Cross-sections are measured in a fiducial phase space defined by the presence of two charged leptons and exactly one jet identified as containing $b$-hadrons.
Six observables are chosen, constructed from the masses and energies of leptons and jets as well as the transverse momenta of neutrinos.
Measurements are normalised with the fiducial cross-section, causing several of the main uncertainties to cancel out.
Dominant uncertainties arise from limited data statistics, signal modelling, and \ttbar{} background modelling.
Results are found to be in good agreement with predictions from several MC event generators.
 
\section*{Acknowledgements}
 
We thank CERN for the very successful operation of the LHC, as well as the
support staff from our institutions without whom ATLAS could not be
operated efficiently.
 
We acknowledge the support of ANPCyT, Argentina; YerPhI, Armenia; ARC, Australia; BMWFW and FWF, Austria; ANAS, Azerbaijan; SSTC, Belarus; CNPq and FAPESP, Brazil; NSERC, NRC and CFI, Canada; CERN; CONICYT, Chile; CAS, MOST and NSFC, China; COLCIENCIAS, Colombia; MSMT CR, MPO CR and VSC CR, Czech Republic; DNRF and DNSRC, Denmark; IN2P3-CNRS, CEA-DRF/IRFU, France; SRNSF, Georgia; BMBF, HGF, and MPG, Germany; GSRT, Greece; RGC, Hong Kong SAR, China; ISF, I-CORE and Benoziyo Center, Israel; INFN, Italy; MEXT and JSPS, Japan; CNRST, Morocco; NWO, Netherlands; RCN, Norway; MNiSW and NCN, Poland; FCT, Portugal; MNE/IFA, Romania; MES of Russia and NRC KI, Russian Federation; JINR; MESTD, Serbia; MSSR, Slovakia; ARRS and MIZ\v{S}, Slovenia; DST/NRF, South Africa; MINECO, Spain; SRC and Wallenberg Foundation, Sweden; SERI, SNSF and Cantons of Bern and Geneva, Switzerland; MOST, Taiwan; TAEK, Turkey; STFC, United Kingdom; DOE and NSF, United States of America. In addition, individual groups and members have received support from BCKDF, the Canada Council, CANARIE, CRC, Compute Canada, FQRNT, and the Ontario Innovation Trust, Canada; EPLANET, ERC, ERDF, FP7, Horizon 2020 and Marie Sk{\l}odowska-Curie Actions, European Union; Investissements d'Avenir Labex and Idex, ANR, R{\'e}gion Auvergne and Fondation Partager le Savoir, France; DFG and AvH Foundation, Germany; Herakleitos, Thales and Aristeia programmes co-financed by EU-ESF and the Greek NSRF; BSF, GIF and Minerva, Israel; BRF, Norway; CERCA Programme Generalitat de Catalunya, Generalitat Valenciana, Spain; the Royal Society and Leverhulme Trust, United Kingdom.
 
The crucial computing support from all WLCG partners is acknowledged gratefully, in particular from CERN, the ATLAS Tier-1 facilities at TRIUMF (Canada), NDGF (Denmark, Norway, Sweden), CC-IN2P3 (France), KIT/GridKA (Germany), INFN-CNAF (Italy), NL-T1 (Netherlands), PIC (Spain), ASGC (Taiwan), RAL (UK) and BNL (USA), the Tier-2 facilities worldwide and large non-WLCG resource providers. Major contributors of computing resources are listed in Ref.~\cite{ATL-GEN-PUB-2016-002}.
 
\printbibliography{}
 
\clearpage
\begin{flushleft}
{\Large The ATLAS Collaboration}

\bigskip

M.~Aaboud$^\textrm{\scriptsize 137d}$,
G.~Aad$^\textrm{\scriptsize 88}$,
B.~Abbott$^\textrm{\scriptsize 115}$,
O.~Abdinov$^\textrm{\scriptsize 12}$$^{,*}$,
B.~Abeloos$^\textrm{\scriptsize 119}$,
S.H.~Abidi$^\textrm{\scriptsize 161}$,
O.S.~AbouZeid$^\textrm{\scriptsize 139}$,
N.L.~Abraham$^\textrm{\scriptsize 151}$,
H.~Abramowicz$^\textrm{\scriptsize 155}$,
H.~Abreu$^\textrm{\scriptsize 154}$
% ...
\bigskip
\\
% ...
$^{12}$ Institute of Physics, Azerbaijan Academy of Sciences, Baku, Azerbaijan\\
% ...
$^{88}$ CPPM, Aix-Marseille Universit{\'e} and CNRS/IN2P3, Marseille, France\\
% ...
$^{115}$ Homer L. Dodge Department of Physics and Astronomy, University of Oklahoma, Norman OK, United States of America\\
% ...
$^{119}$ LAL, Univ. Paris-Sud, CNRS/IN2P3, Universit{\'e} Paris-Saclay, Orsay, France\\
% ...
$^{137}$ $^{(a)}$ Facult{\'e} des Sciences Ain Chock, R{\'e}seau Universitaire de Physique des Hautes Energies - Universit{\'e} Hassan II, Casablanca; $^{(b)}$ Centre National de l'Energie des Sciences Techniques Nucleaires, Rabat; $^{(c)}$ Facult{\'e} des Sciences Semlalia, Universit{\'e} Cadi Ayyad, LPHEA-Marrakech; $^{(d)}$ Facult{\'e} des Sciences, Universit{\'e} Mohamed Premier and LPTPM, Oujda; $^{(e)}$ Facult{\'e} des sciences, Universit{\'e} Mohammed V, Rabat, Morocco\\
% ...
$^{139}$ Santa Cruz Institute for Particle Physics, University of California Santa Cruz, Santa Cruz CA, United States of America\\
% ...
$^{151}$ Department of Physics and Astronomy, University of Sussex, Brighton, United Kingdom\\
% ...
$^{154}$ Department of Physics, Technion: Israel Institute of Technology, Haifa, Israel\\
$^{155}$ Raymond and Beverly Sackler School of Physics and Astronomy, Tel Aviv University, Tel Aviv, Israel\\
% ...
$^{161}$ Department of Physics, University of Toronto, Toronto ON, Canada\\
% ...
$^{a}$ Also at Department of Physics, King's College London, London, United Kingdom\\
$^{b}$ Also at Institute of Physics, Azerbaijan Academy of Sciences, Baku, Azerbaijan\\
$^{c}$ Also at Novosibirsk State University, Novosibirsk, Russia\\
$^{d}$ Also at TRIUMF, Vancouver BC, Canada\\
$^{e}$ Also at Department of Physics {\&} Astronomy, University of Louisville, Louisville, KY, United States of America\\
% ...
$^{*}$ Deceased
\end{flushleft}

 
\end{document}